\RequirePackage{fixltx2e}
\documentclass[12pt,a4paper]{iopart}
\pdfoutput=1
\usepackage[utf8]{inputenc}
\usepackage[unicode=true,bookmarksopen=true,colorlinks=true,urlcolor=blue,anchorcolor=blue,citecolor=blue,filecolor=blue,linkcolor=blue,menucolor=blue,pagecolor=blue,linktocpage=true,pdfa=true]{hyperref}

\usepackage{graphicx}

\makeatletter
\providecommand*\@nameundef[1]{\expandafter\let\csname #1\endcsname\@undefined}
\@nameundef{equation*}
\@nameundef{endequation*}
\makeatother
\usepackage{amsmath}
\usepackage{amssymb}
\usepackage{amsfonts}

\usepackage{mathtools}
\usepackage{lmodern}
\usepackage[T1]{fontenc}
\usepackage{bbm}
\DeclareMathAlphabet{\mathbfi}{OML}{cmm}{b}{it}

\usepackage{subcaption}

\let\originalleft\left
\let\originalright\right
\renewcommand{\left}{\mathopen{}\mathclose\bgroup\originalleft}
\renewcommand{\right}{\aftergroup\egroup\originalright}

\makeatletter
\newenvironment{equations}[1][]{\subequations\ifx\relax#1\relax\else\label{#1}\fi\align\ignorespaces}{\endalign\ignorespacesafterend\endsubequations}
\def\@spliteq#1{\begin{equation}\begin{split}#1\end{split}\end{equation}}
\def\splitequation{\collect@body\@spliteq}

\makeatother

\renewcommand{\vec}[1]{{\ifnum9<1#1\mathbf{#1}\else\ifcat\noexpand#1\relax\boldsymbol{#1}\else\mathbfi{#1}\fi\fi}}
\newcommand{\mathe}{\mathrm{e}}
\newcommand{\mathi}{\mathrm{i}}
\let\oldre\Re
\let\oldim\Im
\renewcommand{\Re}{\oldre\mathfrak{e}\,}
\renewcommand{\Im}{\oldim\mathfrak{m}\,}
\newcommand{\total}{\mathop{}\!\mathrm{d}}
\newcommand{\laplace}{\mathop{}\!\bigtriangleup}

\newcommand{\eqend}[1]{\,#1}
\newcommand{\bigo}[1]{\mathcal{O}\left({#1}\right)}

\newcommand{\expect}[1]{\left\langle{#1}\right\rangle}

\newcommand{\xx}{\text{\scriptsize$\mathcal{X}$}}

\usepackage[numbers,sort&compress]{natbib}
\allowdisplaybreaks

\begin{document}

\title{Gauge-invariant quantum gravitational corrections to correlation functions}

\author{Markus B. Fr{\"o}b}
\address{Department of Mathematics, University of York, Heslington, York, YO10 5DD, United Kingdom}

\ead{mbf503@york.ac.uk}

\begin{abstract}
A recent proposal for gauge-invariant observables in inflation [R. Brunetti et al., \href{http://dx.doi.org/10.1007/JHEP08(2016)032}{JHEP \textbf{1608} (2016) 032}] is examined. We give a generalisation of their construction to general background spacetimes. In flat space, we calculate one-loop graviton corrections to a scalar two-point function in a general gauge for the graviton. We explicitely show how the gauge-dependent terms cancel between the usual self-energy contributions and the additional corrections inherent in these observables. The one-loop corrections have the expected functional form, contrary to another recently studied proposal for gauge-invariant observables [M. B. Fr{\"o}b, \href{http://dx.doi.org/10.1088/1361-6382/aa9ad1}{Class.~Quant.~Grav. \textbf{35} (2018) 035005}] where this is not the case. Furthermore, we determine the one-loop graviton corrections to the four-point coupling of the gauge-invariant scalar field, and the corresponding running of the coupling constant induced by graviton loops. Interestingly, the $\beta$ function is negative for most values of the non-minimal coupling of the scalar field to curvature, including minimal and conformal coupling.

\end{abstract}

\submitto{CQG}

\maketitle

\section{Introduction}
\label{sec_introduction}

While the study of theories of quantum gravity has a long history, the only observational evidence for quantum gravity to date comes from cosmology, namely the fluctuations of the cosmic microwave background~\cite{planck2015a,planck2015b,planck2015c}. They are well described by considering linear metric fluctuations around a classical background, i.e., in the perturbative approach to quantum gravity. Even though perturbative quantum gravity is not power-counting renormalisable, one can treat it in the sense of an effective field theory~\cite{burgess2004}, and obtain unambiguous predictions at scales below the Planck scale. However, if one wants to generalise the treatment to the non-linear level (i.e., including graviton loops), a serious difficulty arises: the construction of gauge-invariant observables. Because of diffeomorphism invariance, it is well known that there can be no local observables in General Relativity (or any other metric theory of gravity) in general. Nevertheless, when one only considers linear perturbations around a fixed background, it is possible to find a complete set of gauge-invariant local observables starting from an IDEAL characterisation of the background spacetime~\cite{canepadappiaggikhavkine2017}. An IDEAL\footnote{Intrinsic, Deductive, Explicit and ALgorithmic, or Rainich-type~\cite{rainich1925}.} characterisation of a given spacetime is a set of tensorial equations $T_a[g,\phi] = 0$ constructed covariantly out of the metric $g_{\mu\nu}$, the curvature tensors and other scalar or tensor fields $\phi$ and their derivatives, which are satisfied if and only if the given spacetime is locally isometric to the reference spacetime. For example, maximally symmetric spaces are characterised by $R_{\mu\nu\rho\sigma} - 2 k g_{\mu[\rho} g_{\sigma]\nu} = 0$ (where $k = 0$ corresponds to Minkowski, $k > 0$ to de Sitter and $k < 0$ to anti-de Sitter spacetime), while the Schwarzschild geometry of mass $m$ is characterised by the conditions~\cite{ferrandosaez1998}
\begin{splitequation}
&R_{\mu\nu} = 0 \eqend{,} \qquad C^{\mu\nu\alpha\beta} C_{\alpha\beta\rho\sigma} - w C^{\mu\nu}{}_{\rho\sigma} - 2 w^2 \delta^\mu_{[\rho} \delta_{\sigma]}^\nu = 0 \eqend{,} \qquad C^{\mu[\nu\rho]\sigma} \nabla_\mu w \nabla_\sigma w = 0 \eqend{,} \\
&w \neq 0 \eqend{,} \qquad \alpha > 0 \eqend{,} \qquad \left( 2 w u^\mu u^\nu C_{\mu\rho\nu\sigma} - w^2 g_{\rho\sigma} + 2 w^2 u_\rho u_\sigma \right) \nabla^\rho w \nabla^\sigma w > 0 \eqend{,}
\end{splitequation}
where
\begin{equation}
w = - \left( \frac{1}{12} C^{\mu\nu\alpha\beta} C_{\alpha\beta\rho\sigma} C^{\rho\sigma}{}_{\mu\nu} \right)^\frac{1}{3} \eqend{,} \qquad \alpha = \frac{1}{9 w^2} \nabla^\mu w \nabla_\mu w - 2 w \eqend{,} \qquad m = w \alpha^{-\frac{3}{2}} \eqend{,}
\end{equation}
and $u^\mu$ is any unit time-like vector. For more general spacetimes, including the Friedmann--Lema{\^\i}tre--Robertson--Walker (FLRW) spacetimes which are used in cosmology, more complicated conditions are needed, which then also involve the inflaton field~\cite{canepadappiaggikhavkine2017}.

Considering now linear perturbations around a background, an infinitesimal coordinate transformation $x^\mu = x^\mu + \delta x^\mu = x^\mu - \xi^\mu$ leads to a gauge transformation of the metric perturbation $h_{\mu\nu}$ of the form
\begin{equation}
\delta h_{\mu\nu} = \nabla_\mu \xi_\nu + \nabla_\nu \xi_\mu \eqend{,}
\end{equation}
and in general the change of the perturbation $A^{(1)}$ of a quantity $\tilde{A} = A^{(0)} + A^{(1)}$ is given by the Lie derivative of the background,
\begin{equation}
\delta A^{(1)} = \mathcal{L}_\xi A^{(0)} \eqend{.}
\end{equation}
Obviously, when the background quantity vanishes, the first-order perturbation is gauge-invariant, which in particular is the case for the IDEAL characterisation tensors $T_a[g,\phi]$. Since furthermore the characterisation is complete, we have $T^{(1)}_a[h,\phi^{(1)}] = 0$ for all $a$ if and only if $h$ and $\phi^{(1)}$ are pure gauge, i.e., of the form $h_{\mu\nu} = \nabla_\mu \xi_\nu + \nabla_\nu \xi_\mu$, $\phi^{(1)} = \mathcal{L}_\xi \phi^{(0)}$ for some $\xi$~\cite{canepadappiaggikhavkine2017}. Therefore, the $T^{(1)}_a[h,\phi^{(1)}]$ provide a complete set of local and gauge-invariant observables in the linearised theory.

This nice construction unfortunately does not extend to higher orders, and one is forced to seek other approaches. One of these are so-called relational observables (see the recent~\cite{rovelli1991,dittrich2006,giddingsmarolfhartle2006,tambornino2012,khavkine2015,marolf2015,brunettifredenhagenrejzner2016} and references therein), which can be loosely defined as the value of some dynamical field $\Phi$ of the theory at the point where some other dynamical field $\Phi'$ has a prescribed value. In general, one takes four scalar fields $\tilde{X}^\mu[\Phi]$ as field-dependent coordinates on which to evaluate the other dynamical fields of the theory. To be able to extract sufficiently local information from the theory, it is of course necessary that the four scalar fields satisfy an appropriate non-degeneracy condition. In the perturbative approach, this condition means that one must be able to distinguish all points of the background spacetime by the background values $X^\mu$ of the $\tilde{X}^\mu$, which becomes problematic if the background spacetime is highly symmetric, such as Minkowski space or the FLRW spacetimes. In these cases, one can add additional scalar fields to the theory (e.g., the Brown--Kucha{\v r} dust~\cite{brownkuchar1995}), but this changes the physical content of theory. A way out of this dilemma has recently been achieved by Brunetti et\,al.~\cite{brunettietal2016} for FLRW spacetimes, with background metric
\begin{equation}
g_{\mu\nu} \total x^\mu \total x^\nu = a(\eta)^2 \left( - \total \eta^2 + \total \vec{x}^2 \right) \eqend{,}
\end{equation}
which is a solution of the Einstein equations with a scalar field $\phi(\eta)$ (the inflaton), where $\eta$ is conformal time, and where the form of the scale factor $a(\eta)$ depends on the scalar potential. (In this case, a complete set of local gauge-invariant observables in the linear theory was constructed in~\cite{froebhackhiguchi2017}.) We will review their solution in section~\ref{sec_construction}, and give a generalisation to more general backgrounds including Minkowski space. In section~\ref{sec_inv2pf} we calculate the invariant scalar two-point function including one-loop gravitational corrections in flat space, and in section~\ref{sec_coupling} we calculate the gravitional corrections to the running $\lambda \phi^4$ coupling. We conclude with an outlook towards further work and open problems in section~\ref{sec_discussion}. Some technical computations are relegated to the appendices. We use the `+++' convention of~\cite{mtw_book}, work mostly in $n$ dimensions, and set $c = \hbar = 1$ and $\kappa^2 = 16 \pi G_\text{N}$.

\section{Construction of gauge-invariant observables}
\label{sec_construction}

The idea of~\cite{brunettietal2016} is based on the observation that in the background spacetime, the spatial coordinates $x^i$ are harmonic, fulfilling the differential equation $\laplace x^i = 0$. Since the Laplace operator $\laplace$ transforms as a scalar under rotations and translations of the spatial submanifold, one can construct the required remaining three scalar functionals order by order in perturbation theory by imposing that they are harmonic with respect to the perturbed Laplacian, and that they reduce to the spatial coordinates for vanishing metric perturbation. An explicit formula can be given as follows: in the perturbed geometry with inflaton $\tilde{\phi}$ and metric $\tilde{g}$, one first defines the unit time-like vector normal to the equal-time hypersurfaces
\begin{equation}
\tilde{n}^\mu \equiv \frac{\tilde{g}^{\mu\nu} \partial_\nu \tilde{\phi}}{\sqrt{ - \tilde{g}^{\mu\nu} \partial_\mu \tilde{\phi} \partial_\nu \tilde{\phi}}}
\end{equation}
and the spatial metric
\begin{equation}
\tilde{\gamma}_{\mu\nu} \equiv \tilde{g}_{\mu\nu} + \tilde{n}_\mu \tilde{n}_\nu \eqend{.}
\end{equation}
The perturbed Laplacian (acting on scalar functions) is then given by
\begin{equation}
\tilde{\laplace}_\phi f \equiv \tilde{\gamma}_\rho^\mu \tilde{\nabla}_\mu \left( \tilde{\gamma}^{\rho\nu} \tilde{\nabla}_\nu f \right) \eqend{,}
\end{equation}
and we define $n-1$ scalar functionals $\tilde{X}^i$ ($i = 1,\ldots,n-1$) by imposing that
\begin{equation}
\label{laplace_tildex}
\tilde{\laplace}_\phi \tilde{X}^i = 0 \eqend{,}
\end{equation}
and that their unperturbed value is equal to the background spatial coordinates: $X^i = x^i$ (we note that the background value of the Laplacian is $\laplace_\phi = a^{-2} \laplace$ with the scale factor $a$). One can, in fact, give en explicit formula to calculate the $\tilde{X}^i$ order by order in perturbation theory. For this, let $\tilde{G}_\phi$ be the Green's function for $\tilde{\laplace}_\phi$ with vanishing boundary conditions at spatial infinity (and $G_\phi$ its background value, the Green's function for $\laplace_\phi$), and denote the difference between perturbed and background value by a $\delta$. We calculate
\begin{equation}
\laplace_\phi G_\phi = \tilde{\laplace}_\phi \tilde{G}_\phi = \left( \laplace_\phi + \delta\! \laplace_\phi \right) \left( G_\phi + \delta G_\phi \right) \eqend{,}
\end{equation}
and from this by convoluting with $G_\phi$ on the left
\begin{equation}
\delta G_\phi = - G_\phi \cdot \delta\! \laplace_\phi \left( G_\phi + \delta G_\phi \right) = G_\phi \cdot \sum_{n=1}^\infty \left( - \delta\! \laplace_\phi G_\phi \cdot \right)^n \eqend{,}
\end{equation}
where the last equality follows by repeatedly replacing $\delta G_\phi$ by the right-hand side. The solution of eq.~\eqref{laplace_tildex} is then given by
\begin{equation}
\label{tildex_explicit_sol}
\tilde{X}^i = \left( 1 - \tilde{G}_\phi \cdot \tilde{\laplace}_\phi \right) x^i = \sum_{n=0}^\infty \left( - G_\phi \cdot \delta\! \laplace_\phi \right)^n x^i \eqend{.}
\end{equation}
Acting with $\tilde{\laplace}_\phi$, this obviously fulfils eq.~\eqref{laplace_tildex}, and since $\tilde{G}_\phi \cdot \tilde{\laplace}_\phi$ is the identity only on functions which vanish at spatial infinity, the solution is not vacuous.

This idea can be generalised in various ways. For example, still in the inflationary context, one could replace the Laplacian by the d'Alembertian $\tilde{\nabla}^2$, since for the background spacetime also $\nabla^2 x^i = a^{-2} \left[ \partial^2 - (n-2) H a \partial_\eta \right] x^i = 0$~\cite{froeblima2017}. This does not work for the time $\eta$, which would still be defined by taking the perturbed inflaton as time coordinate. However, since inflationary backgrounds are conformally flat, we have
\begin{equation}
\left( \nabla^2 - \xi_\text{cc} R \right) \left( a^{-\frac{n-2}{2}} x^\mu \right) = a^{-\frac{n+2}{2}} \partial^2 x^\mu = 0
\end{equation}
with the conformal coupling $\xi_\text{cc} = (n-2)/[4(n-1)]$. It is thus possible to define $n$ scalar functionals $\tilde{X}^\mu$, $\mu = 1,\ldots,n$, by imposing
\begin{equation}
\left( \tilde{\nabla}^2 - \xi_\text{cc} \tilde{R} \right) \left[ a^{-\frac{n-2}{2}}(\tilde{X}) \tilde{X}^\mu \right] = 0 \eqend{,}
\end{equation}
where $a(\tilde{X})$ is defined by replacing $\eta$ by $\tilde{X}^0$ in the scale factor (i.e., keeping the functional form). Finally, perturbing around flat space one can simply impose
\begin{equation}
\label{flat_space_perturbed_harmonic}
\tilde{\nabla}^2 \tilde{X}^\mu = 0
\end{equation}
as a direct generalisation of the background relation $\partial^2 x^\mu = 0$. In all these cases, we have an explicit solution for the $\tilde{X}^\mu$, given by the obvious generalisation of the formula~\eqref{tildex_explicit_sol}. However, while for the Laplacian (an elliptic operator) there is a unique Green's function, for the hyperbolic operator $\tilde{\nabla}^2$ there are many Green's functions to choose from, and the concrete choice forms part of the definition of invariant observables. We will see later on that the Feynman propagator is the correct choice in our case.

In fact, the latter condition can be used for more general spacetimes. In contrast to flat space, in general the background coordinates $x^\mu$ will not be harmonic, but they satisfy the equation
\begin{equation}
\nabla^2 x^\mu = \frac{1}{\sqrt{-g}} \partial_\rho \left( \sqrt{-g} \, g^{\mu\rho} \right) = - g^{\rho\sigma} \Gamma^\mu_{\rho\sigma} \eqend{,}
\end{equation}
where $\nabla^2$ is the scalar d'Alembertian of the background metric $g_{\mu\nu}$. One therefore has to pass to the new coordinates
\begin{equation}
y^\mu(x) = x^\mu + \int G(x,x') \left( \sqrt{-g} g^{\rho\sigma} \Gamma^\mu_{\rho\sigma} \right)(x') f(x') \total^n x' \eqend{,}
\end{equation}
where $G(x,x')$ is any Green's function of the scalar d'Alembertian,
\begin{equation}
\nabla^2 G(x,x') = \frac{\delta^n(x-x')}{\sqrt{-g}} \eqend{,}
\end{equation}
and $f$ is a cutoff function (smooth and with compact support). The existence of such a Green's function with either retarded or advanced boundary conditions is guaranteed for globally hyperbolic spacetimes (see, e.g.,~\cite{baerginouxpfaeffle2007}), and since $f$ has compact support the integral is finite as well. Since
\begin{equation}
\nabla^2 y^\mu(x) = \left( g^{\rho\sigma} \Gamma^\mu_{\rho\sigma} \right)(x) \left[ 1 - f(x) \right] \eqend{,}
\end{equation}
the new coordinates are harmonic at all points where $f = 1$. Whether harmonic coordinates exist globally (i.e., if it is possible to set $f = 1$ in the whole spacetime) is a more difficult problem; we assume that they exist for the spacetime under consideration. Perturbing the metric in these new coordinates, we can then again impose the condition~\eqref{flat_space_perturbed_harmonic} to determine the field-dependent coordinates $\tilde{X}^{(\mu)}$.

As explained in the introduction, invariant observables are obtained by evaluating the dynamical fields of the theory on these field-dependent coordinates. That is, we invert the relation between the background coordinates $x^\mu$ and the $\tilde{X}^\mu$ to the desired order in perturbation theory, and perform a coordinate transformation from the background coordinates $x^\mu$ to the $\tilde{X}^\mu$. For a scalar field $\phi$, the corresponding invariant, which we denote by $\phi(\xx)$, is just given by the scalar field $\phi(x)$ where $\tilde{X}$ is held fixed, while for higher-spin fields one also has to include the Jacobian of the transformation.

\subsection{Second-order perturbative expansion in flat space}
\label{sec_construction_pert}

We use $\kappa$ as a perturbative parameter, and expand
\begin{equation}
\tilde{X}^\alpha = \sum_{k=0}^\infty \kappa^k X_{(k)}^\alpha \eqend{,}
\end{equation}
with $X_{(0)}^\alpha = x^\alpha$. To second order in $\kappa$, the expansion of the perturbed metric and its determinant is given by
\begin{equations}[construction_metric_expansion]
\tilde{g}_{\mu\nu} &= \eta_{\mu\nu} + \kappa h_{\mu\nu} \eqend{,} \\
\tilde{g}^{\mu\nu} &= \eta^{\mu\nu} - \kappa h^{\mu\nu} + \kappa^2 h^{\mu\rho} h_\rho^\nu + \bigo{\kappa^3} \eqend{,} \\
\sqrt{-\tilde{g}} &= 1 + \frac{1}{2} \kappa h + \frac{1}{8} \kappa^2 h^2 - \frac{1}{4} \kappa^2 h^{\mu\nu} h_{\mu\nu} + \bigo{\kappa^3} \eqend{.}
\end{equations}
The condition that we impose on the $\tilde{X}^\alpha$ is
\begin{equation}
0 = \tilde{\nabla}^2 \tilde{X}^\alpha = \frac{1}{\sqrt{-\tilde{g}}} \partial_\mu \left( \sqrt{-\tilde{g}} \, \tilde{g}^{\mu\nu} \partial_\nu \tilde{X}^\alpha \right) \eqend{,}
\end{equation}
and to second order in $\kappa$ we obtain
\begin{equations}[construction_pert_d2x]
\partial^2 X_{(1)}^\alpha &= J_{(1)}^\alpha[h] \eqend{,} \\
\partial^2 X_{(2)}^\alpha &= J_{(2)}^\alpha[h] + K_{(1)}[h] X_{(1)}^\alpha
\end{equations}
with
\begin{equations}
J_{(1)}^\alpha &= \partial_\mu h^{\mu\alpha} - \frac{1}{2} \partial^\alpha h \eqend{,} \\
J_{(2)}^\alpha &= - \partial^\mu \left( h_{\mu\nu} h^{\nu\alpha} \right) + \frac{1}{2} h^{\alpha\mu} \partial_\mu h + \frac{1}{2} h_{\mu\nu} \partial^\alpha h^{\mu\nu} \eqend{,} \\
K_{(1)} &= h^{\mu\nu} \partial_\mu \partial_\nu + \left( \partial_\mu h^{\mu\nu} - \frac{1}{2} \partial^\nu h \right) \partial_\nu \eqend{.}
\end{equations}
Given a Green's function $G(x,y)$ which fulfils
\begin{equation}
\label{construction_pert_green}
\partial^2 G(x,y) = \delta^n(x-y) \eqend{,}
\end{equation}
the solution of eqns.~\eqref{construction_pert_d2x} is given by
\begin{equations}[construction_pert_xresult]
X_{(1)}^\alpha(x) &= \int G(x,y) J_{(1)}^\alpha(y) \total^n y \eqend{,} \\
X_{(2)}^\alpha(x) &= \int G(x,y) \left[ J_{(2)}^\alpha(y) + K_{(1)}(y) X_{(1)}^\alpha(y) \right] \total^n y \eqend{.}
\end{equations}

It is now easy to check that the $\tilde{X}^\alpha$ transform as scalars, using that
\begin{equation}
\label{hmunu_gauge_trafo}
\delta_\xi h_{\mu\nu} = 2 \partial_{(\mu} \xi_{\nu)} + \kappa \xi^\rho \partial_\rho h_{\mu\nu} + 2 \kappa h_{\rho(\mu} \partial_{\nu)} \xi^\rho \eqend{.}
\end{equation}
We calculate
\begin{equations}
\delta_\xi J_{(1)}^\alpha &= \partial^2 \xi^\alpha + \kappa \left( \xi^\nu \partial_\nu J_{(1)}^\alpha + J_{(1)}^\nu \partial^\alpha \xi_\nu + 2 \partial_{(\mu} \xi_{\nu)} \partial^\mu h^{\nu\alpha} + \partial^2 \xi_\nu h^{\nu\alpha} - \partial_\mu \xi_\nu \partial^\alpha h^{\mu\nu} \right) \eqend{,} \\
\delta_\xi J_{(2)}^\alpha &= - 2 \partial_{(\mu} \xi_{\nu)} \partial^\mu h^{\nu\alpha} - \partial^2 \xi_\nu h^{\nu\alpha} + \partial_\mu \xi_\nu \partial^\alpha h^{\mu\nu} - 2 J^{(1)}_\nu \partial^{(\alpha} \xi^{\nu)} - h_{\mu\nu} \partial^\mu \partial^\nu \xi^\alpha + \bigo{\kappa} \eqend{,} \\
\delta_\xi K_{(1)} &= 2 \left( \partial^\mu \xi^\nu \right) \partial_\mu \partial_\nu + \left( \partial^2 \xi^\nu \right) \partial_\nu + \bigo{\kappa} \eqend{,}
\end{equations}
and from this, integrating by parts where necessary, and using that $G(x,y)$ is a Green's function~\eqref{construction_pert_green},
\begin{equations}[construction_xi_gaugetrafo]
\begin{split}
\delta_\xi X_{(1)}^\alpha(x) &= \xi^\alpha(x) + \kappa \xi_\nu(x) h^{\nu\alpha}(x) \\
&\quad+ \kappa \int \xi_\nu(y) \left[ \partial^y_\mu G(x,y) \left( \partial^\mu h^{\nu\alpha} - \partial^\nu h^{\mu\alpha} + \partial^\alpha h^{\mu\nu} \right)(y) - \partial_y^\alpha G(x,y) J_{(1)}^\nu(y) \right] \total^n y \eqend{,}
\end{split} \\
\begin{split}
\delta_\xi X_{(2)}^\alpha(x) &= \xi^\nu(x) \partial_\nu X_{(1)}^\alpha(x) - \xi_\nu(x) h^{\nu\alpha}(x) + \bigo{\kappa} \\
&\quad- \int \xi_\nu(y) \left[ \partial^y_\mu G(x,y) \left( \partial^\mu h^{\nu\alpha} - \partial^\nu h^{\mu\alpha} + \partial^\alpha h^{\mu\nu} \right)(y) - \partial_y^\alpha G(x,y) J_{(1)}^\nu(y) \right] \total^n y \eqend{.}
\end{split}
\end{equations}
It follows that up to second order in $\kappa$
\begin{splitequation}
\label{construction_tildex_gaugetrafo}
\delta_\xi \tilde{X}^\alpha &= \kappa \delta_\xi X_{(1)}^\alpha + \kappa^2 \delta_\xi X_{(2)}^\alpha + \bigo{\kappa^3} \\
&= \kappa \xi^\alpha + \kappa^2 \xi^\nu \partial_\nu X_{(1)}^\alpha + \bigo{\kappa^3} = \kappa \xi^\nu \partial_\nu \tilde{X}^\alpha \eqend{,}
\end{splitequation}
and as postulated, the $\tilde{X}^\alpha$ do transform as scalars.

To construct the invariant scalar field, we first invert the relation between $x$ and $\tilde{X}$. To second order in $\kappa$, we obtain
\begin{splitequation}
x^\alpha &= \tilde{X}^\alpha - \kappa X_{(1)}^\alpha(x) - \kappa^2 X_{(2)}^\alpha(x) + \bigo{\kappa^3} \\
&= \tilde{X}^\alpha - \kappa X_{(1)}^\alpha\left[ \tilde{X} - \kappa X_{(1)}(\tilde{X}) \right] - \kappa^2 X_{(2)}^\alpha(\tilde{X}) + \bigo{\kappa^3} \\
&= \tilde{X}^\alpha - \kappa X_{(1)}^\alpha(\tilde{X}) + \kappa^2 X_{(1)}^\mu(\tilde{X}) \partial_\mu X_{(1)}^\alpha(\tilde{X}) - \kappa^2 X_{(2)}^\alpha(\tilde{X}) + \bigo{\kappa^3} \eqend{.}
\end{splitequation}
The invariant observable, which we denote by $\phi(\xx)$, is obtained by evaluating the scalar field $\phi$ at the point $x$ (holding $\tilde{X}$ fixed), which gives
\begin{splitequation}
\label{construction_invscalar_def}
\phi(\xx) \equiv \phi(x) &= \phi - \kappa X_{(1)}^\alpha \partial_\alpha \phi \\
&\quad- \kappa^2 X_{(2)}^\alpha \partial_\alpha \phi + \kappa^2 X_{(1)}^\mu \partial_\mu X_{(1)}^\alpha \partial_\alpha \phi + \frac{1}{2} \kappa^2 X_{(1)}^\alpha X_{(1)}^\beta \partial_\alpha \partial_\beta \phi + \bigo{\kappa^3} \eqend{,}
\end{splitequation}
where all terms in the second line are evaluated at $\tilde{X}$. Using that
\begin{equation}
\delta_\xi \phi = \kappa \xi^\mu \partial_\mu \phi
\end{equation}
and the change~\eqref{construction_xi_gaugetrafo} or~\eqref{construction_tildex_gaugetrafo} of the $X_{(i)}^\alpha$ under a gauge transformation, it is straightforward to check that
\begin{equation}
\delta_\xi \phi(\xx) = \bigo{\kappa^3} \eqend{.}
\end{equation}
We note that in the generalised Landau (exact) gauge $\partial_\mu h^{\mu\alpha} = \frac{1}{2} \partial^\alpha h$, we have $J_{(1)}^\alpha = 0$ and thus $X_{(1)}^\alpha = 0$. For the non-linear Landau gauge
\begin{equation}
\label{landau_nonlinear}
J_{(1)}^\alpha + \kappa J_{(2)}^\alpha = \partial_\mu h^{\mu\alpha} - \frac{1}{2} \partial^\alpha h + \kappa \left[ - \partial^\mu \left( h_{\mu\nu} h^{\nu\alpha} \right) + \frac{1}{2} h^{\alpha\mu} \partial_\mu h + \frac{1}{2} h_{\mu\nu} \partial^\alpha h^{\mu\nu} \right] = 0 \eqend{,}
\end{equation}
we even have $X_{(1)}^\alpha + \kappa X_{(2)}^\alpha = \bigo{\kappa^2}$, and thus the coordinate corrections to the scalar observable $\phi$ vanish in this gauge, to the order we are working.

\section{The invariant scalar two-point function}
\label{sec_inv2pf}

\subsection{Perturbative expansion to one-loop order}

While the above construction of invariant observables is valid for the perturbative quantisation of any (metric) theory of gravity, to calculate correlation functions of the invariant scalar~\eqref{construction_invscalar_def} we obviously need to choose a theory, that is, an action. We take the usual Einstein--Hilbert action
\begin{equation}
\label{grav_action}
S_\text{G} = \frac{1}{\kappa^2} \int \tilde{R} \sqrt{-\tilde{g}} \total^n x
\end{equation}
for gravity, and a non-minimally coupled action for the scalar field
\begin{equation}
\label{matter_action}
S_\text{M} = - \frac{1}{2} \int \left( \tilde{g}^{\mu\nu} \partial_\mu \phi \partial_\nu \phi + m^2 \phi^2 + \xi \tilde{R} \phi^2 \right) \sqrt{-\tilde{g}} \total^n x \eqend{.}
\end{equation}
To obtain the one-loop corrections, we expand this action to second order in metric perturbations around flat space, which results in
\begin{equation}
S_\text{M} = S_\text{M}^{(0)} + \kappa S_\text{M}^{(1)} + \bigo{\kappa^2}
\end{equation}
for the matter action, with
\begin{equations}
S_\text{M}^{(0)} &= - \frac{1}{2} \int \left( \partial^\mu \phi \partial_\mu \phi + m^2 \phi^2 \right) \total^n x \eqend{,} \label{matter_action_0} \\
S_\text{M}^{(1)} &= \frac{1}{2} \int h^{\mu\nu} \left( \tau_{\mu\nu\rho\sigma} \partial^\rho \phi \partial^\sigma \phi - P_{\mu\nu} \phi^2 \right) \total^n x \eqend{,}
\end{equations}
and where we defined the tensor
\begin{equation}
\tau_{\mu\nu\rho\sigma} \equiv \eta_{\rho(\mu} \eta_{\nu)\sigma} - \frac{1}{2} \eta_{\mu\nu} \eta_{\rho\sigma}
\end{equation}
and the symmetric differential operator
\begin{equation}
P_{\mu\nu} \equiv \frac{1}{2} \eta_{\mu\nu} m^2 + \xi \left( \partial_\mu \partial_\nu - \eta_{\mu\nu} \partial^2 \right) \eqend{.}
\end{equation}
Note that we have not given an explicit expression for the second-order contributions. This is because to one-loop order, they only contribute to the invariant scalar two-point function through graviton tadpoles, which vanish in dimensional regularisation around flat space. For the same reason, we also only need the tree-order contributions from the gravitational action~\eqref{grav_action}, which is
\begin{equation}
\label{grav_action_0}
S_\text{G}^{(0)} = \frac{1}{4} \int \left( - \partial^\rho h^{\mu\nu} \partial_\rho h_{\mu\nu} + 2 \partial^\mu h_{\mu\rho} \partial_\nu h^{\nu\rho} - 2 \partial_\mu h^{\mu\nu} \partial_\nu h + \partial^\rho h \partial_\rho h \right) \total^n x \eqend{.}
\end{equation}
As is well known, this action is invariant under the gauge transformation~\eqref{hmunu_gauge_trafo} (to zeroth order in $\kappa$), and to fix the gauge invariance we add the gauge-fixing term
\begin{equation}
S_\text{GF}^{(0)} = - \frac{1}{2 \alpha} \int \left[ \partial^\nu h_{\mu\nu} - \left( 1 + \frac{1}{\beta} \right) \partial_\mu h \right] \left[ \partial_\rho h^{\mu\rho} - \left( 1 + \frac{1}{\beta} \right) \partial^\mu h \right] \total^n x \eqend{.}
\end{equation}
The corresponding ghost action also does not contribute to the invariant scalar two-point function at one-loop order since is are no direct coupling between the ghost fields and the scalar. To determine the graviton propagator, we write
\begin{equation}
S_\text{G}^{(0)} + S_\text{GF}^{(0)} = \frac{1}{2} \int h_{\mu\nu} P^{\mu\nu\rho\sigma} h_{\rho\sigma} \total^n x
\end{equation}
with the symmetric differential operator
\begin{splitequation}
P^{\mu\nu\rho\sigma} &\equiv \frac{1}{2} \eta^{\mu(\rho} \eta^{\sigma)\nu} \partial^2 - \left( 1 - \frac{1}{\alpha} \right) \partial^{(\mu} \eta^{\nu)(\rho} \partial^{\sigma)} \\
&\quad+ \left( \frac{1}{2} - \frac{1 + \beta}{\alpha \beta} \right) \left( \eta^{\mu\nu} \partial^\rho \partial^\sigma + \eta^{\rho\sigma} \partial^\mu \partial^\nu \right) - \left( \frac{1}{2} - \frac{(1+\beta)^2}{\alpha \beta^2} \right) \eta^{\mu\nu} \eta^{\rho\sigma} \partial^2 \eqend{.}
\end{splitequation}
The graviton propagator $G_{\mu\nu\rho\sigma}(x,x')$ satisfies
\begin{equation}
P^{\alpha\beta\mu\nu} G_{\mu\nu\rho\sigma}(x,x') = \delta^\alpha_{(\rho} \delta^\beta_{\sigma)} \delta^n(x-x') \eqend{,}
\end{equation}
and it is straightforward to check that
\begin{splitequation}
\label{prop_graviton}
G_{\mu\nu\rho\sigma}(x,x') &= \left( 2 \eta_{\mu(\rho} \eta_{\sigma)\nu} - \frac{2}{n-2} \eta_{\mu\nu} \eta_{\rho\sigma} \right) G_0(x,x') + 4 (\alpha-1) \frac{\partial_{(\mu} \eta_{\nu)(\rho} \partial_{\sigma)}}{\partial^2} G_0(x,x') \\
&\quad+ \frac{2}{n-2} (2+\beta) \left( \eta_{\mu\nu} \frac{\partial_\rho \partial_\sigma}{\partial^2} + \eta_{\rho\sigma} \frac{\partial_\mu \partial_\nu}{\partial^2} \right) G_0(x,x') \\
&\quad- (2+\beta) \left[ \frac{n}{n-2} (2+\beta) + (\alpha-1) (2-\beta) \right] \frac{\partial_\mu \partial_\nu \partial_\rho \partial_\sigma}{\left( \partial^2 \right)^2} G_0(x,x') \eqend{,}
\end{splitequation}
where the scalar propagator $G_{m^2}$ satisfies
\begin{equation}
\label{prop_scalar}
\left( \partial^2 - m^2 \right) G_{m^2}(x,x') = \delta^n(x-x') \eqend{.}
\end{equation}
For later use, we write this in the form
\begin{equation}
G_{\mu\nu\rho\sigma}(x,x') = \sum_{k=1}^5 g^{(k)} T^{(k)}_{\mu\nu\rho\sigma}(\partial) G_0(x,x')
\end{equation}
with
\begin{equations}[gravprop_tensor_structure]
T^{(1)}_{\mu\nu\rho\sigma} &= 2 \eta_{\mu(\rho} \eta_{\sigma)\nu} \eqend{,} \\
T^{(2)}_{\mu\nu\rho\sigma} &= \eta_{\mu\nu} \eta_{\rho\sigma} \eqend{,} \\
T^{(3)}_{\mu\nu\rho\sigma} &= 4 \frac{\partial_{(\mu} \eta_{\nu)(\rho} \partial_{\sigma)}}{\partial^2} \eqend{,} \\
T^{(4)}_{\mu\nu\rho\sigma} &= \eta_{\mu\nu} \frac{\partial_\rho \partial_\sigma}{\partial^2} + \eta_{\rho\sigma} \frac{\partial_\mu \partial_\nu}{\partial^2} \eqend{,} \\
T^{(5)}_{\mu\nu\rho\sigma} &= \frac{\partial_\mu \partial_\nu \partial_\rho \partial_\sigma}{\left( \partial^2 \right)^2}
\end{equations}
and
\begin{splitequation}
g^{(1)} &= 1 \eqend{,} \qquad g^{(2)} = - \frac{2}{n-2} \eqend{,} \qquad g^{(3)} = (\alpha-1) \eqend{,} \qquad g^{(4)} = \frac{2}{n-2} (2+\beta) \eqend{,} \\
g^{(5)} &= - (2+\beta) \left[ \frac{n}{n-2} (2+\beta) + (\alpha-1) (2-\beta) \right] \eqend{.}
\end{splitequation}
Since massless tadpoles vanish in dimensional regularisation, of the non-linear generalised Landau gauge~\eqref{landau_nonlinear} in which the coordinate corrections vanish, only the linear part contributes to one-loop order. This gauge is obtained for $\alpha = 0$ and $\beta = -2$, where the propagator reads
\begin{equation}
\label{propagator_landau}
G_{\mu\nu\rho\sigma}(x,x') = \left[ 2 \eta_{\mu(\rho} \eta_{\sigma)\nu} - \frac{2}{n-2} \eta_{\mu\nu} \eta_{\rho\sigma} - 4 \frac{\partial_{(\mu} \eta_{\nu)(\rho} \partial_{\sigma)}}{\partial^2} \right] G_0(x,x') \eqend{,}
\end{equation}
and fulfils
\begin{equation}
\partial^\mu G_{\mu\nu\rho\sigma}(x,x') = \frac{1}{2} \partial_\nu \eta^{\alpha\beta} G_{\alpha\beta\rho\sigma}(x,x') \eqend{.}
\end{equation}
However, to show explicitly the gauge invariance of the result we will work in this section in the general gauge~\eqref{prop_graviton}.

\begin{figure}[ht]
\includegraphics[width=0.22\textwidth]{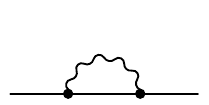}\hfil
\includegraphics[width=0.22\textwidth]{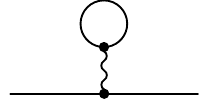}\hfil
\includegraphics[width=0.22\textwidth]{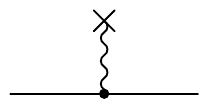}\hfil
\includegraphics[width=0.22\textwidth]{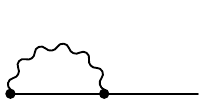}\hfil
\includegraphics[width=0.22\textwidth]{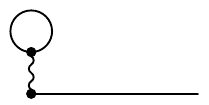}\hfil
\includegraphics[width=0.22\textwidth]{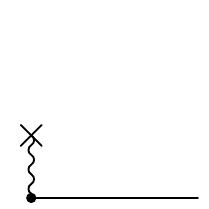}\hfil
\includegraphics[width=0.22\textwidth]{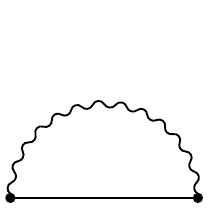}\hfil
\includegraphics[width=0.22\textwidth]{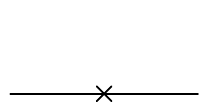}\hfil
\caption{One-loop corrections to the invariant scalar two-point function at order $\kappa^2$. Wiggly lines are gravitons, plain lines are scalars. The first three and the last diagram are the usual corrections to the scalar self-energy (with external propagators), while the middle four diagrams represent coordinate corrections.}
\label{fig_feynman}
\end{figure}

The last terms that we need to add to the action are counterterms. To renormalise the invariant two-point function, we need a wave function and mass counterterm as well as a higher-derivative term, which arises because gravity is perturbatively non-renormalisable:
\begin{equation}
S_\text{CT} = - \frac{1}{2} \int \left( \delta_Z \partial^\mu \phi \partial_\mu \phi + \delta_m \phi^2 \right) \total^n x - \frac{1}{2} \delta_{Z_1} \int \left( \partial^2 \phi \right)^2 \total^n x \eqend{.}
\end{equation}
Furthermore, we need a cosmological-constant counterterm
\begin{equation}
- \frac{2 \delta_\Lambda}{\kappa^2} \int \sqrt{-g} \total^n x \eqend{,}
\end{equation}
which makes a contribution
\begin{equation}
S_\text{CC}^{(1)} = - \frac{\delta_\Lambda}{\kappa^2} \int h \total^n x
\end{equation}
at linear order. The counterterms $\delta_Z$, $\delta_m$, $\delta_{Z_1}$ and $\delta_\Lambda$ are all of order $\kappa^2$. Using the expansion~\eqref{construction_invscalar_def}, we can then calculate the perturbative expansion of the invariant two-point function. Again, because massless tadpoles vanish in dimensional regularisation, we can drop terms containing $X_{(2)}^\alpha(X)$ which gives a graviton tadpole anchored at $X$, and we can also drop all terms which contain an odd number of gravitons. This gives
\begin{splitequation}
\label{inv2pf_exp_ct}
\expect{ \phi(\xx) \phi(\xx') } &= \expect{ \phi(X) \phi(X') }_0 + \mathi \left( \expect{ \phi(X) \phi(X') S_\text{CT} }_0 - \expect{ \phi(X) \phi(X') }_0 \expect{ S_\text{CT} }_0 \right) \\
&\quad- \frac{1}{2} \kappa^2 \left[ \expect{ \phi(X) \phi(X') S^{(1)} S^{(1)} }_0 - \expect{ \phi(X) \phi(X') }_0 \expect{ S^{(1)} S^{(1)} }_0 \right] \\
&\quad- \mathi \kappa^2 \expect{ \phi(X) X_{(1)}^\mu(X') \partial_\mu \phi(X') S^{(1)} }_0 - \mathi \kappa^2 \expect{ X_{(1)}^\alpha(X) \partial_\alpha \phi(X) \phi(X') S^{(1)} }_0 \\
&\quad+ \kappa^2 \expect{ X_{(1)}^\alpha(X) \partial_\alpha \phi(X) X_{(1)}^\mu(X') \partial_\mu \phi(X') }_0 + \bigo{\kappa^4} \eqend{,}
\end{splitequation}
where
\begin{equation}
S^{(1)} \equiv S_\text{M}^{(1)} + S_\text{CC}^{(1)} \eqend{,}
\end{equation}
and $\expect{ \cdot }_0$ is the expectation value in the free theory with the propagators~\eqref{prop_scalar} and~\eqref{prop_graviton}. We can now evaluate the invariant two-point function~\eqref{inv2pf_exp_ct} using Wick's theorem; the corresponding Feynman diagrams are shown in figure~\ref{fig_feynman}.

\subsection{Calculation}

Denoting the invariant two-point function by
\begin{equation}
\mathi \mathcal{G}_{m^2}(\tilde{X},\tilde{X}') \equiv \expect{ \phi(\xx) \phi(\xx') }
\end{equation}
and passing to Fourier space where the scalar propagator~\eqref{prop_scalar} reads
\begin{equation}
G_{m^2}(x,x') = - \int \frac{1}{p^2 + m^2 - \mathi 0} \mathe^{\mathi p (x-x')} \frac{\total^n p}{(2\pi)^n} \eqend{,}
\end{equation}
we obtain
\begin{equation}
\label{inv2pf_calg}
\tilde{\mathcal{G}}_{m^2}(p) = - \frac{1}{p^2 + m^2 - \mathi 0} + \frac{\delta_m + \delta_Z p^2 + \delta_{Z_1} (p^2)^2}{\left( p^2 + m^2 - \mathi 0 \right)^2} + \kappa^2 \sum_{N=1}^7 \sum_{k=1}^5 g^{(k)} I^{(N,k)}(p) \eqend{,}
\end{equation}
where $I^{(N,k)}$ is the contribution from the $k$-th tensor structure of the list~\eqref{gravprop_tensor_structure} to the $N$-th Feynman diagram of figure~\ref{fig_feynman}. They are given by
\begin{equation}
I^{(1,k)} = \frac{\mathi}{4} \frac{1}{( p^2 + m^2 - \mathi 0 )^2} \int \frac{1}{q^2 - \mathi 0} \frac{1}{(p-q)^2 + m^2 - \mathi 0} \mathcal{I}^{(1,k)}(p,q) \frac{\total^n q}{(2\pi)^n}
\end{equation}
with
\begin{equations}[inv2pf_cali1]
\begin{split}
\mathcal{I}^{(1,1)}(p,q) &= 4 p^2 (p-q)^2 + 2 (n-2) [ p^2 - (pq) ] [ p^2 - (pq) + 2 m^2 ] + 2 n m^4 \\
&\quad- 16 \xi [ p^2 q^2 - (pq)^2 ] + 8 (n-1) \xi q^2 [ p^2 - (pq) + m^2 + \xi q^2 ] \eqend{,}
\end{split} \\
\mathcal{I}^{(1,2)}(p,q) &= \left[ (n-2) p^2 - (n-2) (pq) + n m^2 + 2 (n-1) \xi q^2 \right]^2 \eqend{,} \\
\mathcal{I}^{(1,3)}(p,q) &= 4 p^2 (p-q)^2 + 8 m^2 \left[ p^2 + (pq) - 2 \frac{(pq)^2}{q^2} \right] + 4 m^4 \eqend{,} \\
\begin{split}
\mathcal{I}^{(1,4)}(p,q) &= 2 \left[ (n-2) p^2 - (n-2) (pq) + n m^2 + 2 (n-1) \xi q^2 \right] \left[ p^2 + (pq) - 2 \frac{(pq)^2}{q^2} + m^2 \right] \eqend{,}
\end{split} \\
\mathcal{I}^{(1,5)}(p,q) &= \left[ p^2 + (pq) - 2 \frac{(pq)^2}{q^2} + m^2 \right]^2 \eqend{,}
\end{equations}
\begin{equation}
I^{(2,k)} = \frac{\mathi}{8} \frac{1}{( p^2 + m^2 - \mathi 0 )^2} \iint \frac{1}{k^2 - \mathi 0} \frac{1}{q^2 + m^2 - \mathi 0} \mathcal{I}^{(2,k)}(p,q,k) \delta^n(k) \total^n k \frac{\total^n q}{(2\pi)^n}
\end{equation}
with
\begin{equations}[inv2pf_cali2]
\mathcal{I}^{(2,1)}(p,q,k) &= 2 (n-2) ( p^2 + m^2 ) ( q^2 + m^2 ) - 4 q^2 p^2 + 8 (pq)^2 + 4 m^4 \eqend{,} \\
\mathcal{I}^{(2,2)}(p,q,k) &= \left[ (n-2) p^2 + n m^2 \right] \left[ (n-2) q^2 + n m^2 \right] \eqend{,} \\
\mathcal{I}^{(2,3)}(p,q,k) &= 16 \frac{(pk) (qk)}{k^2} \left[ (pq) - \frac{(pk) (qk)}{k^2} \right] + 4 \left[ p^2 + m^2 - 2 \frac{(pk)^2}{k^2} \right] \left[ q^2 + m^2 - 2 \frac{(qk)^2}{k^2} \right] \eqend{,} \\
\begin{split}
\mathcal{I}^{(2,4)}(p,q,k) &= \left[ (n-2) p^2 + n m^2 \right] \left[ q^2 + m^2 - 2 \frac{(qk)^2}{k^2} \right] \\
&\quad+ \left[ (n-2) q^2 + n m^2 \right] \left[ p^2 + m^2 - 2 \frac{(pk)^2}{k^2} \right] \eqend{,}
\end{split} \\
\mathcal{I}^{(2,5)}(p,q,k) &= \left[ p^2 + m^2 - 2 \frac{(pk)^2}{k^2} \right] \left[ q^2 + m^2 - 2 \frac{(qk)^2}{k^2} \right] \eqend{,}
\end{equations}
\begin{equation}
\label{inv2pf_i3}
I^{(3,k)} = - \frac{\delta_\Lambda}{2 \kappa^2} \frac{1}{( p^2 + m^2 - \mathi 0 )^2} \int \frac{1}{k^2 - \mathi 0} \mathcal{I}^{(3,k)}(p,k) \delta^n(k) \total^n k
\end{equation}
with
\begin{equations}[inv2pf_cali3]
\mathcal{I}^{(3,1)}(p,k) &= 2 (n-2) p^2 + 2 n m^2 \eqend{,} \\
\mathcal{I}^{(3,2)}(p,k) &= n (n-2) p^2 + n^2 m^2 \eqend{,} \\
\mathcal{I}^{(3,3)}(p,k) &= - 8 \frac{(pk)^2}{k^2} + 4 (p^2+m^2) \eqend{,} \\
\mathcal{I}^{(3,4)}(p,k) &= 2 (n-1) p^2 + 2 n m^2 - 2 n \frac{(pk)^2}{k^2} \eqend{,} \\
\mathcal{I}^{(3,5)}(p,k) &= - 2 \frac{(pk)^2}{k^2} + (p^2+m^2) \eqend{,}
\end{equations}
\begin{equation}
\label{inv2pf_i4}
I^{(4,k)} = - \frac{\mathi}{4} \frac{1}{p^2 + m^2 - \mathi 0} \int \left[ \tilde{G}(q) + \tilde{G}(-q) \right] \frac{1}{q^2 - \mathi 0} \frac{1}{(p-q)^2 + m^2 - \mathi 0} \mathcal{I}^{(4,k)}(p,q) \frac{\total^n q}{(2\pi)^n}
\end{equation}
with
\begin{equations}[inv2pf_cali4]
\begin{split}
\mathcal{I}^{(4,1)}(p,q) &= - 4 (pq) (p-q)^2 - 2 (n-2) [ (pq) - q^2 ] [ p^2 - (pq) + m^2 ] \\
&\qquad- 4 (n-1) \xi q^2 [ (pq) - q^2 ] \eqend{,}
\end{split} \\
\mathcal{I}^{(4,2)}(p,q) &= - (n-2) [ (pq) - q^2 ] [ (n-2) p^2 - (n-2) (pq) + n m^2 + 2 (n-1) \xi q^2 ] \eqend{,} \\
\mathcal{I}^{(4,3)}(p,q) &= - 4 (pq) (p-q)^2 + 4 [ (pq) - q^2 ] m^2 \eqend{,} \\
\mathcal{I}^{(4,4)}(p,q) &= 2 [ (pq) - q^2 ] \left[ (n-2) \frac{(pq)}{q^2} [ (pq) - q^2 ] + m^2 + (n-1) \xi q^2 \right] \eqend{,} \\
\mathcal{I}^{(4,5)}(p,q) &= [ (pq) - q^2 ] \left[ p^2 + (pq) - 2 \frac{(pq)^2}{q^2} + m^2 \right] \eqend{,}
\end{equations}
\begin{splitequation}
\label{inv2pf_i5}
I^{(5,k)} = - \frac{\mathi}{8} \frac{1}{p^2 + m^2 - \mathi 0} &\iint \frac{\tilde{G}(q)}{q^2 - \mathi 0} \frac{1}{k^2 + m^2 - \mathi 0} \mathcal{I}^{(5,k)}(p,q,k) \frac{\total^n k}{(2\pi)^n} \\
&\quad\times \left( \mathe^{\mathi q X} - \mathe^{\mathi q X'} \right) \delta^n(q) \total^n q
\end{splitequation}
with
\begin{equations}[inv2pf_cali5]
\mathcal{I}^{(5,1)}(p,q,k) &= - 8 (pk) (qk) + 4 (pq) k^2 - 2 (n-2) (pq) ( k^2 + m^2 ) \eqend{,} \\
\mathcal{I}^{(5,2)}(p,q,k) &= - (n-2) (pq) [ n ( k^2 + m^2 ) - 2 k^2 ] \eqend{,} \\
\mathcal{I}^{(5,3)}(p,q,k) &= - 8 (pk) (qk) + 4 (pq) ( k^2 + m^2 ) \eqend{,} \\
\mathcal{I}^{(5,4)}(p,q,k) &= 2 (pq) \left[ (n-2) \frac{(qk)^2}{q^2} + m^2 \right] \eqend{,} \\
\mathcal{I}^{(5,5)}(p,q,k) &= (pq) \left[ k^2 + m^2 - 2 \frac{(qk)^2}{q^2} \right] \eqend{,}
\end{equations}
\begin{equation}
\label{inv2pf_i6}
I^{(6,k)} = \frac{\delta_\Lambda}{2 \kappa^2} \frac{1}{p^2 + m^2 - \mathi 0} \int \frac{\tilde{G}(q)}{q^2 - \mathi 0} \mathcal{I}^{(6,k)}(p,q) \left( \mathe^{\mathi q X} - \mathe^{\mathi q X'} \right) \delta^n(q) \total^n q
\end{equation}
with
\begin{equations}[inv2pf_cali6]
\mathcal{I}^{(6,1)}(p,q) &= - 2 (n-2) (pq) \eqend{,} \\
\mathcal{I}^{(6,2)}(p,q) &= - n (n-2) (pq) \eqend{,} \\
\mathcal{I}^{(6,3)}(p,q) &= 4 (pq) \eqend{,} \\
\mathcal{I}^{(6,4)}(p,q) &= 2 (pq) \eqend{,} \\
\mathcal{I}^{(6,5)}(p,q) &= (pq) \eqend{,}
\end{equations}
and
\begin{equation}
\label{inv2pf_i7}
I^{(7,k)} = \frac{\mathi}{4} \int \tilde{G}(q) \tilde{G}(-q) \frac{1}{q^2 - \mathi 0} \frac{1}{(p-q)^2 + m^2 - \mathi 0} \mathcal{I}^{(7,k)}(p,q) \frac{\total^n q}{(2\pi)^n}
\end{equation}
with
\begin{equations}[inv2pf_cali7]
\mathcal{I}^{(7,1)}(p,q) &= 4 q^2 (p-q)^2 + 2 (n-2) [ (pq) - q^2 ]^2 \eqend{,} \\
\mathcal{I}^{(7,2)}(p,q) &= (n-2)^2 [ (pq) - q^2 ]^2 \eqend{,} \\
\mathcal{I}^{(7,3)}(p,q) &= 4 q^2 (p-q)^2 \eqend{,} \\
\mathcal{I}^{(7,4)}(p,q) &= - 2 (n-2) [ (pq) - q^2 ]^2 \eqend{,} \\
\mathcal{I}^{(7,5)}(p,q) &= [ (pq) - q^2 ]^2 \eqend{.}
\end{equations}

Let us first consider the second and third diagrams, $I^{(2,k)}$ and $I^{(3,k)}$. While the closed massive scalar loop (the integral over $q$) is well-defined, it is attached to a zero-momentum massless graviton (the integral over $k$), which leads to terms of the form $k^{-2} \delta^n(k)$ that are clearly infinite even in dimensional regularisation. Such tadpoles also appear in quantum corrections to the graviton propagator~\cite{capper1975}, where they are cancelled by the cosmological-constant counterterm. We can cancel them in a similar way, performing the $q$ integral in $I^{(2,k)}$ first. Since it is rotation invariant, we use
\begin{equation}
\label{inv2pf_rotinv_int}
\int q^\mu q^\nu f(q^2) \frac{\total^n q}{(2\pi)^n} = \frac{\eta^{\mu\nu}}{n} \int q^2 f(q^2) \frac{\total^n q}{(2\pi)^n}
\end{equation}
to reduce the tensor structure and obtain purely scalar integrals. Because scaleless integrals vanish in dimensional regularisation, we further obtain
\begin{equation}
\int \frac{q^2}{q^2 + m^2 - \mathi 0} \frac{\total^n q}{(2\pi)^n} = - m^2 \int \frac{1}{q^2 + m^2 - \mathi 0} \frac{\total^n q}{(2\pi)^n} \eqend{,}
\end{equation}
and it follows that
\begin{equation}
\label{inv2pf_i2_decomp}
I^{(2,k)} = \frac{\mathi}{4 n} m^2 \frac{1}{( p^2 + m^2 - \mathi 0 )^2} \int \frac{1}{k^2 - \mathi 0} \mathcal{I}^{(3,k)}(p,k) \delta^n(k) \total^n k \int \frac{1}{q^2 + m^2 - \mathi 0} \frac{\total^n q}{(2\pi)^n} \eqend{.}
\end{equation}
Comparing with equation~\eqref{inv2pf_i3}, the contribution of equation~\eqref{inv2pf_i2_decomp} can be cancelled by the counterterm
\begin{equation}
\delta_\Lambda = \frac{\mathi}{2 n} \kappa^2 m^2 \int \frac{1}{q^2 + m^2 - \mathi 0} \frac{\total^n q}{(2\pi)^n} = \frac{\Gamma\left( - \frac{n}{2} \right)}{4 (4\pi)^\frac{n}{2}} \kappa^2 (m^2)^\frac{n}{2}
\end{equation}
[using the standard loop integral~\eqref{app_integrals_tadpole}], which is exactly the result of~\cite{capper1975}, taking into account the different normalisation. In the same way (and with the same $\delta_\Lambda$), we obtain
\begin{equation}
I^{(5,k)} + I^{(6,k)} = 0 \eqend{.}
\end{equation}

The remaining diagrams could now be calculated using the standard tensor reduction and loop integrals given in~\ref{app_integrals}. However, since we are calculating an invariant correlation function, many terms will cancel, and this cancellation can already be seen at the level of the integrands. We thus calculate
\begin{splitequation}
\label{inv2pf_termsumint}
I^{(1,k)} + I^{(4,k)} + I^{(7,k)} &= \frac{\mathi}{4} \frac{1}{( p^2 + m^2 - \mathi 0 )^2} \int \frac{1}{q^2 - \mathi 0} \frac{1}{(p-q)^2 + m^2 - \mathi 0} \\
&\quad\times \Big[ \mathcal{I}^{(1,k)}(p,q) - ( p^2 + m^2 ) \left[ \tilde{G}(q) + \tilde{G}(-q) \right] \mathcal{I}^{(4,k)}(p,q) \\
&\qquad\qquad\qquad+ ( p^2 + m^2 )^2 \tilde{G}(q) \tilde{G}(-q) \mathcal{I}^{(7,k)}(p,q) \Big] \frac{\total^n q}{(2\pi)^n} \eqend{,}
\end{splitequation}
and the expression in brackets reduces to
\begin{splitequation}
&4 \frac{p^2}{q^2 - \mathi 0} [ (p-q)^2 + m^2 ]^2 - 4 ( p^2 + m^2 ) \left[ \tilde{G}(q) + \tilde{G}(-q) + \frac{2}{q^2 - \mathi 0} \right] [ p^2 - (pq) ] [ (p-q)^2 + m^2 ] \\
&\quad+ 4 ( p^2 + m^2 )^2 \left[ \tilde{G}(q) + \frac{1}{q^2 - \mathi 0} \right] \left[ \tilde{G}(-q) + \frac{1}{q^2 - \mathi 0} \right] q^2 (p-q)^2 \\
\end{splitequation}
for $k = 3$,
\begin{splitequation}
&2 \frac{(pq)}{q^2 - \mathi 0} [ (p-q)^2 + m^2 ] \left[ 2 (n-2) p^2 + 2 (n-1) \left( m^2 + \xi q^2 \right) - (n-2) \frac{(pq)}{q^2} ( p^2 + q^2 + m^2 ) \right] \\
&\quad- 2 ( p^2 + m^2 ) \left[ \tilde{G}(q) + \tilde{G}(-q) + \frac{2}{q^2 - \mathi 0} \right] \left[ \frac{(pq)}{q^2 - \mathi 0} - 1 \right] \\
&\qquad\qquad\times \left[ - (n-2) [ (pq) - q^2 ] ( p^2 + m^2 - (pq) ) + m^2 q^2 + (n-1) \xi (q^2)^2 \right] \\
&\quad- 2 (n-2) ( p^2 + m^2 )^2 \left[ \tilde{G}(q) + \frac{1}{q^2 - \mathi 0} \right] \left[ \tilde{G}(-q) + \frac{1}{q^2 - \mathi 0} \right] [ (pq) - q^2 ]^2
\end{splitequation}
for $k = 4$, and
\begin{splitequation}
&\frac{(pq)^2}{( q^2 - \mathi 0 )^2} [ (p-q)^2 + m^2 ]^2 + ( p^2 + m^2 )^2 \left[ \tilde{G}(q) + \frac{1}{q^2 - \mathi 0} \right] \left[ \tilde{G}(-q) + \frac{1}{q^2 - \mathi 0} \right] [ (pq) - q^2 ]^2 \\
&\quad- ( p^2 + m^2 ) \left[ \tilde{G}(q) + \tilde{G}(-q) + \frac{2}{q^2 - \mathi 0} \right] \frac{(pq)}{q^2 - \mathi 0} [ (pq) - q^2 ] [ (p-q)^2 + m^2 ]
\end{splitequation}
for $k = 5$. We see that if we take the Feynman propagator for the so far unspecified Green's function $G$~\eqref{construction_pert_green} that appears in the construction of the perturbed harmonic coordinates, we have $\tilde{G}(q) = - 1/(q^2 - \mathi 0)$ and most of the terms vanish. The remaining ones contain a factor of $[ (p-q)^2 + m^2 ]$, which cancels the corresponding factor in the integral~\eqref{inv2pf_termsumint} and leaves behind a scaleless $q$-integral. Since these vanish in dimensional regularisation, we have $I^{(1,k)} + I^{(4,k)} + I^{(7,k)} = 0$ for $k = 3$, $4$ and $5$, corresponding to the gauge-dependent terms of the graviton propagator. With this choice of $G$, and replacing
\begin{equation}
(pq) \to \frac{1}{2} ( m^2 + p^2 + q^2 )
\end{equation}
since the difference between left- and right-hand side results in a scaleless integral, we obtain for the graviton loop
\begin{splitequation}
\label{inv2pf_termsumint_grav}
&g^{(1)} I^{(1,1)} + g^{(2)} I^{(1,2)} = \mathi \frac{1}{( p^2 + m^2 - \mathi 0 )^2} \int \frac{1}{q^2 - \mathi 0} \frac{1}{(p-q)^2 + m^2 - \mathi 0} \\
&\qquad\times \bigg[ - 2 p^2 m^2 - \frac{2}{n-2} m^4 + \xi ( p^2 + m^2 )^2 + [ m^2 - 2 \xi ( p^2 - m^2 ) ] q^2 \\
&\qquad\qquad- \frac{4 (n-1)}{n-2} \xi m^2 q^2 + \xi ( q^2 )^2 - \frac{2 (n-1)}{n-2} \xi^2 (q^2)^2 \bigg] \frac{\total^n q}{(2\pi)^n} \eqend{,}
\end{splitequation}
and for the coordinate corrections
\begin{equation}
\label{inv2pf_termsumint_coord}
\sum_{k=1}^2 g^{(k)} \left( I^{(4,k)} + I^{(7,k)} \right) = \mathi m^2 \int \frac{1}{[ q^2 - \mathi 0 ]^2} \frac{1}{(p-q)^2 + m^2 - \mathi 0} \frac{\total^n q}{(2\pi)^n} \eqend{.}
\end{equation}
The integrals are done in~\ref{app_integrals}, and summing we obtain for the graviton loop
\begin{splitequation}
\label{inv2pf_termsumint_grav2}
&(4\pi)^2 \left[ g^{(1)} I^{(1,1)} + g^{(2)} I^{(1,2)} \right] = \frac{\xi ( p^2 )^2}{( p^2 + m^2 - \mathi 0 )^2} \left[ \frac{2}{n-4} + \gamma - \ln (4\pi) + \ln m^2 - 2 \right] \\
&\quad- \frac{m^2 p^2}{( p^2 + m^2 - \mathi 0 )^2} \left[ \left( 2 - 3 \xi - 3 \xi^2 \right) \left[ \frac{2}{n-4} + \gamma - \ln (4\pi) + \ln m^2 \right] + \left( - 4 + 5 \xi + 4 \xi^2 \right) \right] \\
&\quad- \frac{m^4}{( p^2 + m^2 - \mathi 0 )^2} \left[ \left( 2 - 6 \xi + 3 \xi^2 \right) \left[ \frac{2}{n-4} + \gamma - \ln (4\pi) + \ln m^2 \right] + \left( - 4 + 9 \xi - 4 \xi^2 \right) \right] \\
&\quad- \left[ \frac{2 p^2 m^2 + m^4}{( p^2 + m^2 - \mathi 0 )^2} - \xi \right] \left( 1 + \frac{m^2}{p^2 - \mathi 0} \right) \ln\left( 1 + \frac{p^2 - \mathi 0}{m^2} \right) + \bigo{n-4} \eqend{.}
\end{splitequation}
Note that for minimal coupling $\xi = 0$ and in the massless limit $m = 0$, this vanishes identically, in accordance with~\cite{kahyawoodard2007}. Also, for $m = 0$ all terms quadratic in the non-minimal coupling $\xi$ vanish, in accordance with~\cite{borankahyapark2014} (taking the flat-space limit of their result). For the coordinate corrections, we obtain
\begin{splitequation}
\label{inv2pf_termsumint_coord2}
&(4\pi)^2 \sum_{k=1}^2 g^{(k)} \left( I^{(4,k)} + I^{(7,k)} \right) = - \frac{m^2}{p^2 + m^2 - \mathi 0} \left[ \frac{2}{n-4} + \gamma - \ln (4\pi) + \ln m^2 \right] \\
&\qquad\qquad- \frac{m^2}{p^2 + m^2 - \mathi 0} \left( 1 - \frac{m^2}{p^2 - \mathi 0} \right) \ln\left( 1 + \frac{p^2 - \mathi 0}{m^2} \right) + \bigo{n-4} \eqend{.}
\end{splitequation}
The terms in the first lines can be absorbed in the counterterms $\delta_m$, $\delta_Z$ and $\delta_{Z_1}$. Comparing with equation~\eqref{inv2pf_calg}, we need to take
\begin{equations}[inv2pf_counterterms]
\delta_{Z_1} &= \frac{\kappa^2}{(4\pi)^2} \bigg[ - \xi \left[ \frac{2}{n-4} + \gamma - \ln (4\pi) + \ln m^2 \right] + 2 \xi \bigg] + \kappa^2 \delta_{Z_1}^\text{fin} \eqend{,} \\
\delta_Z &= \frac{\kappa^2}{(4\pi)^2} m^2 \bigg[ 3 \left( 1 - \xi - \xi^2 \right) \left[ \frac{2}{n-4} + \gamma - \ln (4\pi) + \ln m^2 \right] - 4 + 5 \xi + 4 \xi^2 \bigg] + \kappa^2 \delta_Z^\text{fin} \eqend{,} \\
\delta_m &= \frac{\kappa^2}{(4\pi)^2} m^4 \bigg[ 3 \left( 1 - \xi \right)^2 \left[ \frac{2}{n-4} + \gamma - \ln (4\pi) + \ln m^2 \right] - 4 + 9 \xi - 4 \xi^2 \bigg] + \kappa^2 \delta_m^\text{fin} \eqend{,}
\end{equations}
and the invariant scalar two-point function including one-loop graviton corrections reads
\begin{splitequation}
\label{inv2pf_result}
\tilde{\mathcal{G}}_{m^2}(p) &= - \frac{1}{p^2 + m^2 - \mathi 0} + \kappa^2 \frac{\delta_m^\text{fin} + \delta_Z^\text{fin} p^2 + \delta_{Z_1}^\text{fin} (p^2)^2}{\left( p^2 + m^2 - \mathi 0 \right)^2} \\
&\quad- \frac{\kappa^2}{(4\pi)^2} \left( \frac{3 m^2}{p^2 + m^2 - \mathi 0} - \xi \frac{p^2 + m^2}{p^2 - \mathi 0} \right) \ln\left( 1 + \frac{p^2 - \mathi 0}{m^2} \right) \eqend{.}
\end{splitequation}
Note that this result is in no particular renormalisation scheme, and that it is seemingly divergent as $m \to 0$ because of the explicit $\ln m^2$ in the counterterms. To obtain a result which is also valid in the massless limit, we would have to impose renormalisation conditions at some fixed, non-vanishing external momenta, which leads to a change of scheme involving $\ln m^2/\mu^2$, where $\mu$ is the renormalisation scale. Alternatively, we can perform the limit first in the regularised expressions~\eqref{inv2pf_termsumint_grav2} and~\eqref{inv2pf_termsumint_coord2} and renormalise afterwards, obtaining
\begin{equation}
\label{inv2pf_result_massless}
\tilde{\mathcal{G}}_0(p) = - \frac{1}{p^2 - \mathi 0} + \kappa^2 \delta_{Z_1}^\text{fin}(\mu) + \xi \frac{\kappa^2}{(4\pi)^2} \ln\left( \frac{p^2 - \mathi 0}{\mu^2} \right)
\end{equation}
with the counterterms
\begin{equations}[massless_higherder_ct]
\delta_{Z_1} &= - \frac{\kappa^2 \xi}{(4\pi)^2} \left[ \frac{2}{n-4} + \gamma - \ln (4\pi) + \ln \mu^2 - 2 \right] + \kappa^2 \delta_{Z_1}^\text{fin}(\mu) \eqend{,} \\
\delta_Z &= \delta_m = 0 \eqend{.}
\end{equations}

It is instructive to see the result also in coordinate space, at least for the massless case. To transform back, we need the integral (\cite{smirnov2004}~Eq.~(A.40), \cite{brychkovprudnikov}~Eq.~(8.715), converted to our conventions)
\begin{equation}
\int \mathe^{\mathi p x} (p^2 - \mathi 0)^\alpha \frac{\total^n p}{(2\pi)^n} = \mathi \frac{4^\alpha \Gamma\left( \frac{n}{2}+\alpha \right)}{\pi^\frac{n}{2} \Gamma(-\alpha)} (x^2 + \mathi 0)^{-\alpha-\frac{n}{2}} \eqend{,}
\end{equation}
valid in $n$ dimensions for $-n/2 < \Re \alpha < 0$. Writing
\begin{equation}
\ln (p^2 - \mathi 0) = p^2 \lim_{\epsilon \to 0^+} \left[ \frac{1}{\epsilon} \left( (p^2 - \mathi 0)^{\epsilon-1} - (p^2 - \mathi 0)^{-1} \right) \right] \eqend{,}
\end{equation}
we obtain
\begin{equations}
\int \frac{\mathe^{\mathi p x}}{p^2 - \mathi 0} \frac{\total^4 p}{(2\pi)^4} &= \frac{\mathi}{4 \pi^2 (x^2 + \mathi 0)} \eqend{,} \\
\int \mathe^{\mathi p x} \ln\left( \frac{p^2 - \mathi 0}{\mu^2} \right) \frac{\total^4 p}{(2\pi)^4} &= \mathi \partial^2 \left[ \frac{\ln\left( \mu^2 (x^2 + \mathi 0) \right)}{4 \pi^2 (x^2 + \mathi 0)} \right] + 2 (\ln 2 - \gamma) \delta^n(x) \eqend{.} \label{inv2pf_fouriertrafo_log}
\end{equations}
It follows that
\begin{equation}
\label{inv2pf_result_massless_coordinate}
\mathi \mathcal{G}_0(x) = \frac{1}{4 \pi^2 (x^2 + \mathi 0)} - \frac{\kappa^2 \xi}{(4\pi)^2} \partial^2 \left[ \frac{\ln\left( \mu^2 (x^2 + \mathi 0) \right)}{4 \pi^2 (x^2 + \mathi 0)} \right] + \mathi \kappa^2 \delta_{Z_1}^\text{fin}(\mu) \delta^n(x) \eqend{,}
\end{equation}
where we performed an additional finite renormalisation
\begin{equation}
\delta_{Z_1}^\text{fin}(\mu) \to \delta_{Z_1}^\text{fin}(\mu) - 2 \xi \frac{\ln 2 - \gamma}{(4\pi)^2}
\end{equation}
to get rid of the extra terms in the Fourier transform~\eqref{inv2pf_fouriertrafo_log}.

In contrast to other approaches to define gauge-invariant correlation functions in perturbative quantum gravity~\cite{froeb2017}, the result~\eqref{inv2pf_result}, \eqref{inv2pf_result_massless_coordinate} has the expected functional form involving only a single power of $\ln x^2$ at one-loop order. In particular, using the two-point function~\eqref{inv2pf_result_massless_coordinate} to quantify quantum gravitational corrections to a scalar interaction potential, the single logarithm gives terms proportional to $\kappa^2 r^{-3}$, which is exactly the form obtained using S-matrix calculations in similar situations (see~\cite{holsteinross2008} and references therein). A double log $\ln^2 x^2$ as found in other approaches~\cite{froeb2017} would give corrections proportional to $\kappa^2 r^{-3} \ln (\mu r)$ to the tree-level $r^{-1}$ potential. While in the case of matter loop corrections to the graviton two-point function such double-log terms are pure gauge and do not make a contribution to the Newton potential since they are coupled to the conserved stress tensor of the point particle~\cite{duff1974,duffliu2000}, in the case of a scalar potential such an argument does not apply, and it is important that only single-log terms appear in the end result. We can finally compare our result~\eqref{inv2pf_result_massless_coordinate} to another recent approach of obtaining invariant corrections~\cite{miaoprokopecwoodard2017}. Their result can also be expressed as a correction to the (amputated) two-point function, and can be obtained from ours~\eqref{inv2pf_result_massless_coordinate} by replacing the overall factor $\xi$ by $-3$. Since they treat a massless, minimally coupled scalar with $\xi = 0$, the result is genuinely different. However, it agrees with the result of~\cite{ratliff2024}, who also treat a massless, minimally coupled scalar.

\section{Gravitationally induced running of the quartic coupling}
\label{sec_coupling}

The study of gravitational corrections to the running of couplings was initiated by Robinson and Wilczek~\cite{robinsonwilczek2006}, but it was pointed out soon after that the result was gauge-dependent~\cite{pietrykowski2006}. Since the observable $\phi(\xx)$ is gauge invariant, the graviton corrections to its quartic coupling are also invariant, and to determine them in principle we would have to construct the effective action for $\phi(\xx)$. This would involve inverting the expansion~\eqref{construction_invscalar_def} up to the desired order, expressing the action in terms of $\phi(\xx)$ and then construct the effective action by a Legendre transformation. However, we have seen in the last section that to order $\kappa^2$ in dimensional regularisation and in the generalised Landau gauge with $\alpha = 0$ and $\beta = -2$ where the graviton propagator is given by~\eqref{propagator_landau}, the coordinate corrections vanish. Therefore, the effective action for $\phi(\chi)$ is given, to order $\kappa^2$, by the effective action for $\phi(x)$ in the gauge~\eqref{propagator_landau}, which simplifies the calculations enormously. Actually, to compare with existing results we will use the gauge with $\beta = -2$ but $\alpha$ arbitrary, where the graviton propagator reads
\begin{equation}
\label{propagator_landau2}
G_{\mu\nu\rho\sigma}(x,x') = \left[ 2 \eta_{\mu(\rho} \eta_{\sigma)\nu} - \frac{2}{n-2} \eta_{\mu\nu} \eta_{\rho\sigma} + 4 (\alpha-1) \frac{\partial_{(\mu} \eta_{\nu)(\rho} \partial_{\sigma)}}{\partial^2} \right] G_0(x,x') \eqend{.}
\end{equation}
The choice $\alpha = 1$ corresponds to Feynman gauge, and the choice $\alpha = 0$, corresponding to generalised Landau gauge, gives the invariant corrections. To the action we need to add an additional $\phi^4$ interaction term and additional counterterms which we write in the form $\lambda S_\lambda$, with
\begin{splitequation}
S_\lambda &\equiv - \frac{1}{4!} \int \phi^4 \sqrt{-g} \total^n x - \frac{1}{4!} \frac{\delta_\lambda}{\lambda} \int \phi^4 \sqrt{-g} \total^n x - \frac{1}{4!} \frac{\delta_{Z_2}}{\lambda} \int \phi^3 \partial^2 \phi \total^n x \\
&= S_\lambda^{(0)} + \kappa S_\lambda^{(1)} + \bigo{\kappa^2} \eqend{,}
\end{splitequation}
where $S_\lambda^{(0)} = S_\lambda[h_{\mu\nu}=0]$,
\begin{equation}
S_\lambda^{(1)} = - \frac{1}{48} \int \phi^4 h \total^n x \eqend{,}
\end{equation}
and $\delta_\lambda = \bigo{\lambda \kappa^2} + \bigo{\lambda^2}$, $\delta_{Z_2} = \bigo{\lambda \kappa^2}$.

The one-loop effective action is just
\begin{splitequation}
\Gamma[\Phi] &= S[\Phi] + \frac{\mathi}{2} \ln \det \left( \frac{\delta^2 S}{(\delta \Phi)^2} \right) + \text{const} \\
&= S[\Phi] + \frac{\mathi}{2} \tr \ln \left[ 1 + \left( \frac{\delta^2 S^{(0)}}{(\delta \Phi)^2} \right)^{-1} \frac{\delta^2 (S-S^{(0)})}{(\delta \Phi)^2} \right] + \text{const} \eqend{,}
\end{splitequation}
where $\Phi = \{ \phi, h_{\mu\nu} \}$, and the quartic coupling can be determined from the four-point function at some fixed renormalisation point. Expanding to first order in $\lambda$ and $\kappa^2$ and writing
\begin{equation}
G^{(0)}_{\Phi\Phi}(x,y) \equiv \left( \frac{\delta^2 S^{(0)}}{\delta \Phi(x) \delta \Phi(y)} \right)^{-1} \eqend{,}
\end{equation}
we then obtain
\begin{splitequation}
\Gamma[\Phi] &= S[\Phi] + \frac{\mathi}{2} \iint G^{(0)}_{\Phi\Phi}(x,y) \frac{\delta^2}{\delta \Phi(x) \delta \Phi(y)} \left[ \kappa S^{(1)} + S_\text{CT} + \lambda S_\lambda^{(0)} + \kappa \lambda S_\lambda^{(1)} \right] \total^n x \total^n y \\
&\quad- \frac{\mathi}{4} \kappa^2 \iiiint G^{(0)}_{\Phi\Phi}(x,y) \frac{\delta^2}{\delta \Phi(y) \delta \Phi(z)} S^{(1)} G^{(0)}_{\Phi\Phi}(z,u) \frac{\delta^2}{\delta \Phi(u) \delta \Phi(x)} S^{(1)} \total^n x \total^n y \total^n z \total^n u \\
&\quad- \frac{\mathi}{2} \iiiint G^{(0)}_{\Phi\Phi}(x,y) \frac{\delta^2}{\delta \Phi(y) \delta \Phi(z)} \left[ \kappa S^{(1)} + S_\text{CT} \right] G^{(0)}_{\Phi\Phi}(z,u) \\
&\qquad\qquad\times \frac{\delta^2}{\delta \Phi(u) \delta \Phi(x)} \left[ \lambda S_\lambda^{(0)} + \kappa \lambda S_\lambda^{(1)} \right] \total^n x \total^n y \total^n z \total^n u \\
&\quad+ \frac{\mathi}{2} \lambda \kappa^2 \iiint\!\!\iiint G^{(0)}_{\Phi\Phi}(x,y) \frac{\delta^2}{\delta \Phi(y) \delta \Phi(z)} S_\lambda^{(0)} G^{(0)}_{\Phi\Phi}(z,u) \frac{\delta^2}{\delta \Phi(u) \delta \Phi(v)} S^{(1)} G^{(0)}_{\Phi\Phi}(v,w) \\
&\qquad\qquad\times \frac{\delta^2}{\delta \Phi(w) \delta \Phi(x)} S^{(1)} \total^n x \total^n y \total^n z \total^n u \total^n v \total^n w + \bigo{\lambda^2} + \bigo{\kappa^3} \eqend{,}
\end{splitequation}
where similarly to the case of the two-point function~\eqref{inv2pf_exp_ct} we dropped $S^{(2)}$ because it only leads to massless tadpoles which vanish in dimensional regularisation. Since we are only interested in the scalar field, we also set the graviton to zero after differentiation. Using that
\begin{equations}
G^{(0)}_{\phi\phi}(x,y) &= G_{m^2}(x,y) \eqend{,} \\
G^{(0)}_{hh}(x,y) &= G_{\mu\nu\alpha\beta}(x,y) \eqend{,}
\end{equations}
it follows that
\begin{equation}
\Gamma[\phi] = \Gamma^{(0)}[\phi] + \Gamma^{(2)}[\phi] + \Gamma^{(4)}[\phi] + \bigo{\lambda^2} + \bigo{\kappa^3} \eqend{,}
\end{equation}
where $\Gamma^{(k)}$ contains $k$ factors of $\phi$. To determine the running of the couplings, we only need the divergent part of the counterterms, which for $\Gamma^{(2)}[\phi]$ we can obtain from the calculation of the invariant two-point function of the last section. For $\Gamma^{(4)}$, we obtain
\begin{splitequation}
\label{eq:effective_action_gamma4}
\Gamma^{(4)}[\phi] &= - \frac{\lambda}{4!} \int \phi^4 \total^n x - \frac{\delta_\lambda}{4!} \int \phi^4 \total^n x - \frac{\delta_{Z_2}}{4!} \int \phi^3 \partial^2 \phi \total^n x \\
&\quad- \frac{\mathi}{12} \lambda \kappa^2 \iint \phi^3(x) \phi(y) J^{(1)}(x,y) \total^n x \total^n y \\
&\quad- \frac{\mathi}{8} \lambda \kappa^2 \iiint \phi(x) \phi^2(y) \phi(z) J^{(2)}(x,y,z) \total^n x \total^n y \total^n z \\
\end{splitequation}
with
\begin{equation}
J^{(1)}(x,y) = \partial^y_\rho \left[ \partial^y_\sigma G_{m^2}(x,y) \tau^{\mu\nu\rho\sigma} G_{\mu\nu\alpha\beta}(y,x) \eta^{\alpha\beta} \right] + G_{m^2}(x,y) P^{\mu\nu} G_{\mu\nu\alpha\beta}(y,x) \eta^{\alpha\beta}
\end{equation}
and
\begin{splitequation}
J^{(2)}(x,y,z) &= 2 \partial^x_\gamma \left[ \partial^x_\delta G_{m^2}(x,y) \partial^z_\rho \left[ \partial^z_\sigma G_{m^2}(y,z) \tau^{\mu\nu\rho\sigma} G_{\mu\nu\alpha\beta}(z,x) \right] \tau^{\alpha\beta\gamma\delta} \right] \\
&\quad+ 2 G_{m^2}(y,z) \partial^x_\gamma \left[ \partial^x_\delta G_{m^2}(x,y) P^{\mu\nu} G_{\mu\nu\alpha\beta}(z,x) \tau^{\alpha\beta\gamma\delta} \right] \\
&\quad+ 2 G_{m^2}(x,y) \partial^z_\rho \left[ \partial^z_\sigma G_{m^2}(y,z) \tau^{\mu\nu\rho\sigma} P^{\alpha\beta} G_{\mu\nu\alpha\beta}(z,x) \right] \\
&\quad+ 2 G_{m^2}(x,y) G_{m^2}(y,z) P^{\mu\nu} P^{\alpha\beta} G_{\mu\nu\alpha\beta}(z,x) \eqend{.}
\end{splitequation}

Passing to Fourier space and performing the tensor algebra, we obtain with the graviton propagator~\eqref{propagator_landau2}
\begin{splitequation}
\tilde{J}^{(1)}(p) = - 2 \int &\left[ (2-\alpha) p^2 - \alpha (pq) - 2 (1-\alpha) \frac{(pq)^2}{q^2} + \frac{2 (n-1)}{n-2} \xi q^2 + \left( \frac{2(n-1)}{n-2} - \alpha \right) m^2 \right] \\
&\qquad\times \frac{1}{(p-q)^2 + m^2 - \mathi 0} \frac{1}{q^2 - \mathi 0} \frac{\total^n q}{(2\pi)^n} \eqend{,}
\end{splitequation}
and
\begin{equation}
J^{(2)}(x,y,z) = \iint \tilde{J}^{(2)}(p,k) \mathe^{\mathi k (x-y)} \mathe^{\mathi p (z-x)} \frac{\total^n p}{(2\pi)^n} \frac{\total^n k}{(2\pi)^n}
\end{equation}
with
\begin{splitequation}
\label{eq:tildej2}
\tilde{J}^{(2)}(p,k) &= \int \bigg[ (1+\alpha) q^2 k^2 - 2 \alpha q^2 p^2 + (1-\alpha) p^2 k^2 + 2 \alpha p^2 (kq) + 2 \alpha q^2 (kq) - 4 \alpha q^2 (pk) \\
&\qquad\quad+ 2 (1-\alpha) (pq) k^2 + 4 \alpha (kq) (pq) - 4 \alpha (kq)^2 - 4 (1-\alpha) (kq) (pk) \\
&\qquad\quad+ 4 m^2 (pq) + 2 (1-\alpha) m^2 [ p^2 - q^2 + 2 (k-q)^2 ] + 2 \left( \frac{2(n-1)}{n-2} - \alpha \right) m^4 \\
&\qquad\quad+ 8 \xi [ p^2 q^2 - (pq)^2 ] + \frac{8 (n-1)}{n-2} \xi m^2 (p-q)^2 + 2 \frac{2 (n-1)}{n-2} \xi^2 [ (p-q)^2 ]^2 \bigg] \\
&\qquad\times \frac{1}{(p-q)^2 - \mathi 0} \frac{1}{q^2 + m^2 - \mathi 0} \frac{1}{(k-q)^2 + m^2 - \mathi 0} \frac{\total^n q}{(2\pi)^n} \\
&\quad- (1-\alpha) \int \left[ ( p^2 - q^2 )^2 ( k^2 + 2 m^2 ) + 4 q^2 [ (p-q)k ]^2 + 4 ( p^2 - q^2 ) (kq) (q-p)k \right] \\
&\qquad\qquad\qquad\times \frac{1}{[ (p-q)^2 - \mathi 0 ]^2} \frac{1}{q^2 + m^2 - \mathi 0} \frac{1}{(k-q)^2 + m^2 - \mathi 0} \frac{\total^n q}{(2\pi)^n} \eqend{,}
\end{splitequation}
up to terms vanishing under the substitution $p \to k-p$, which is equivalent to the exchange of $x$ and $z$ in $J^{(2)}(x,y,z)$, and which do not contribute to the effective action $\Gamma^{(4)}[\phi]$~\eqref{eq:effective_action_gamma4}.

The $q$ integral in $\tilde{J}^{(1)}(p)$ can easily be done in the same way as in the last section, and we obtain
\begin{equation}
\tilde{J}^{(1)}(p) = \frac{\mathi}{(4\pi)^2} \frac{2}{n-4} m^{n-4} \left[ 2 m^2 \left( 3 - \alpha - 3 \xi \right) + p^2 (3-2\alpha) \right] + \bigo{(n-4)^0} \eqend{,}
\end{equation}
and thus
\begin{splitequation}
J^{(1)}(x,y) = \frac{\mathi}{(4\pi)^2} \frac{2}{n-4} m^{n-4} \bigg[ &- (3-2\alpha) \left( \partial^2 - m^2 \right) \delta^n(x-y) \\
&\quad+ 3 (1-2\xi) m^2 \delta^n(x-y) \bigg] + \bigo{(n-4)^0} \eqend{.}
\end{splitequation}
To obtain a full result for $\tilde{J}^{(2)}(p,k)$ would be much harder, but since we are only interested in the divergent contribution we can simplify its evaluation. Power counting reveals that the only possible divergences comes from the region of large $q$, and dropping all terms which are subdominant in this limit (including terms from expanding the numerators), replacing rotationally invariant terms according to equation~\eqref{inv2pf_rotinv_int} and setting $n = 4$, we have to evaluate
\begin{splitequation}
\tilde{J}^{(2)}(p,k) &\approx \bigg[ 2 \alpha [ k^2 - p^2 - (pk) ] + 6 \xi (p^2 + 2 m^2) + 6 \xi^2 [ p^2 - m^2 - (pk) ] \bigg] \\
&\qquad\times \int \frac{1}{q^2 - \mathi 0} \frac{1}{q^2 + m^2 - \mathi 0} \frac{\total^n q}{(2\pi)^n} + 6 \xi^2 \int \frac{1}{q^2 + m^2 - \mathi 0} \frac{\total^n q}{(2\pi)^n} \eqend{.}
\end{splitequation}
Using the integrals~\eqref{app_integrals_tadpole} and~\eqref{app_integrals_convol1b} from the appendix and reversing the Fourier transformation, we obtain
\begin{splitequation}
J^{(2)}(x,y,z) &\approx \frac{\mathi}{(4\pi)^2} \frac{2}{n-4} m^{n-4} \bigg[ 12 \xi (\xi-1) m^2 \delta^n(x-y) \delta^n(x-z) \\
&\qquad+ ( 2 \alpha + 6 \xi^2 ) \partial^\mu \delta^n(x-y) \partial_\mu \delta^n(x-z) + 2 \alpha \delta^n(x-z) \partial^2 \delta^n(x-y) \\
&\qquad- ( 2 \alpha - 6 \xi - 6 \xi^2 ) \delta^n(x-y) \partial^2 \delta^n(x-z) \bigg] + \bigo{(n-4)^0} \eqend{.}
\end{splitequation}
It follows that
\begin{splitequation}
\Gamma^{(4)}[\phi] &= - \frac{\lambda}{4!} \int \phi^4 \total^n x - \frac{1}{4!} \left[ \delta_\lambda - \frac{2}{n-4} m^{n-4} \frac{\lambda \kappa^2}{(4\pi)^2} 4 m^2 \left( 3 - \alpha - 12 \xi + 9 \xi^2 \right) \right] \int \phi^4 \total^n x \\
&\quad- \frac{1}{4!} \left[ \delta_{Z_2} + \frac{2}{n-4} m^{n-4} \frac{\lambda \kappa^2}{(4\pi)^2} 2 \left( 3 - \alpha - 9 \xi - 3 \xi^2 \right) \right] \int \phi^3 \partial^2 \phi \total^n x + \bigo{(n-4)^0} \eqend{,}
\end{splitequation}
and to make the effective action finite we need the counterterms
\begin{equations}[coupling_counter1]
\delta_\lambda &= \frac{2}{n-4} \mu^{n-4} \frac{\lambda \kappa^2}{(4\pi)^2} 4 m^2 \left( 3 - 12 \xi + 9 \xi^2 - \alpha \right) \eqend{,} \\
\delta_{Z_2} &= - \frac{2}{n-4} \mu^{n-4} \frac{\lambda \kappa^2}{(4\pi)^2} 2 \left( 3 - 9 \xi - 3 \xi^2 - \alpha \right) + \kappa^2 \delta_{Z_2}^\text{fin}
\end{equations}
with the renormalisation scale $\mu$. To determine the counterterms $\delta_m$, $\delta_Z$ and $\delta_{Z_1}$ for the propagator~\eqref{propagator_landau2}, we note that the sum~\eqref{inv2pf_termsumint_grav2} corresponds to Feynman gauge $\alpha = 1$, while the sum of~\eqref{inv2pf_termsumint_grav2} and~\eqref{inv2pf_termsumint_coord2} corresponds to the invariant corrections with $\alpha = 0$. We thus obtain
\begin{equations}[coupling_counter2]
\delta_m &= \frac{2}{n-4} \mu^{n-4} \frac{\kappa^2}{(4\pi)^2} m^4 \left[ 3 (1-\xi)^2 - \alpha \right] + \bigo{(n-4)^0} \eqend{,} \\
\delta_Z &= \frac{2}{n-4} \mu^{n-4} \frac{\kappa^2}{(4\pi)^2} m^2 \left[ 3 (1-\xi-\xi^2) - \alpha \right] + \bigo{(n-4)^0} \eqend{,} \\
\delta_{Z_1} &= - \frac{2}{n-4} \mu^{n-4} \frac{\kappa^2}{(4\pi)^2} \xi + \bigo{(n-4)^0} \eqend{,}
\end{equations}
which for $\alpha = 0$ of course reduce to the previous result~\eqref{inv2pf_counterterms}. For $\alpha = 1$ (Feynman gauge), we also recover the results of~\cite{schuster2008,rodigastschuster2010,rodigast2012} (who work with a minimally coupled scalar field, $\xi = 0$) and the ``conventional results'' of~\cite{mackay2010}, taking into account the different normalisation. However, our results are different from the ``invariant results'' of~\cite{mackay2010}, mainly because their gauge condition in this case necessarily involves the scalar field.

The bare field, mass and coupling $\phi_0$, $m_0^2$ and $\lambda_0$ are related to the renormalised ones by
\begin{equation}
\phi_0 = \sqrt{1+\delta_Z} \, \phi \eqend{,} \qquad m_0^2 = \frac{m^2 + \delta_m}{1 + \delta_Z} \eqend{,} \qquad \lambda_0 = \frac{\lambda + \delta_\lambda}{(1 + \delta_Z)^2} \eqend{,}
\end{equation}
and we can determine the $\mu$ dependence of $\phi$, $m^2$ and $\lambda$ by differentiating, using that the bare parameters are $\mu$-independent. We obtain the $\beta$ and $\gamma$ functions
\begin{equations}
\beta &= \mu \frac{\partial}{\partial \mu} \lambda = - \mu \frac{\partial}{\partial \mu} \delta_\lambda + 2 \frac{\lambda + \delta_\lambda}{1+\delta_Z} \mu \frac{\partial}{\partial \mu} \delta_Z \eqend{,} \\
\gamma &= - \mu \frac{\partial}{\partial \mu} \ln \phi = \frac{1}{2 (1+\delta_Z)} \mu \frac{\partial}{\partial \mu} \delta_Z \eqend{,} \\
\beta_m &= \mu \frac{\partial}{\partial \mu} \frac{m^2}{\mu^2} = \frac{m^2 + \delta_m}{\mu^2 (1 + \delta_Z)} \mu \frac{\partial}{\partial \mu} \delta_Z - \frac{1}{\mu} \frac{\partial}{\partial \mu} \delta_m - 2 \frac{m^2}{\mu^2} \eqend{.}
\end{equations}
Using that from equations~\eqref{coupling_counter1} and~\eqref{coupling_counter2} we have
\begin{equations}
\mu \frac{\partial}{\partial \mu} \delta_\lambda &= 8 \frac{\lambda \kappa^2}{(4\pi)^2} m^2 \left[ 3 - 12 \xi + 9 \xi^2 - \alpha \right] \eqend{,} \\
\mu \frac{\partial}{\partial \mu} \delta_Z &= 2 \frac{\kappa^2}{(4\pi)^2} m^2 \left[ 3 - 3 \xi - 3 \xi^2 - \alpha \right] \eqend{,} \\
\mu \frac{\partial}{\partial \mu} \delta_m &= 2 \frac{\kappa^2}{(4\pi)^2} m^4 \left[ 3 - 6 \xi + 3 \xi^2 - \alpha \right] \eqend{,}
\end{equations}
and neglecting the $\mu$-dependence of $\kappa^2$, at one-loop order we obtain
\begin{equations}[scalar_beta_functions]
\beta &= - \frac{4 \lambda \kappa^2}{(4\pi)^2} m^2 \left[ 3 - 21 \xi + 21 \xi^2 - \alpha \right] + \bigo{\kappa^4} + \bigo{\lambda^2} \eqend{,} \\
\gamma &= \frac{\kappa^2}{(4\pi)^2} m^2 \left[ 3 - 3 \xi - 3 \xi^2 - \alpha \right] + \bigo{\kappa^4} + \bigo{\lambda} \eqend{,} \\
\beta_m &= \frac{m^2}{\mu^2} \left[ 6 \frac{\kappa^2}{(4\pi)^2} m^2 \xi (1 - 2 \xi) - 2 \right] + \bigo{\kappa^4} + \bigo{\lambda} \eqend{.}
\end{equations}
Even in Feynman gauge ($\alpha = 1$), the $\beta$ function is negative for a large range of values of the non-minmal coupling, namely for all $\xi \lesssim 0.11$ or $\xi \gtrsim 0.89$. While this would exclude the conformal coupling $\xi = 1/6$, this result is gauge-dependent. If one considers the invariant corrections ($\alpha = 0$), which lead to a gauge-independent $\beta$ function, it is negative for all $\xi < 1/2 - \sqrt{21}/14 \approx 0.17$ and $\xi > 1/2 + \sqrt{21}/14 \approx 0.83$, which includes the conformal coupling. In fact, the gauge-invariant $\beta$ function attains its maximum value at $\xi = 1/2$, where (reinstating Newton's constant)
\begin{equation}
\beta_{1/2} = \frac{9 \lambda \kappa^2}{(4\pi)^2} m^2 \approx 2.865 \lambda G_\text{N} m^2 \eqend{,}
\end{equation}
and for minimal ($\xi = 0$) and conformal coupling we obtain
\begin{equation}
\beta_\text{min} \approx - 3.820 \lambda G_\text{N} m^2 \eqend{,} \qquad \beta_\text{conf} \approx - 0.106 \lambda G_\text{N} m^2 \eqend{.}
\end{equation}

We can determine the running of the couplings of the higher-derivative terms in the same way. The relation between the bare counterterms $\delta_{Z_{1/2}}^0$ and the renormalised ones is given by
\begin{equation}
\delta_{Z_1}^0 = \frac{\delta_{Z_1}}{1 + \delta_Z} \eqend{,} \qquad \delta_{Z_2}^0 = \frac{\delta_{Z_2}}{(1 + \delta_Z)^2} \eqend{.}
\end{equation}
Using that the bare counterterms are independent of the renormalisation scale $\mu$, we calculate from the explicit results~\eqref{coupling_counter1} and~\eqref{coupling_counter2} that
\begin{equations}
\mu \frac{\partial}{\partial \mu} \delta_{Z_1}^\text{fin} &= \frac{2 \xi}{(4\pi)^2} + \bigo{\kappa^2} + \bigo{\lambda} \eqend{,} \\
\mu \frac{\partial}{\partial \mu} \delta_{Z_2}^\text{fin} &= 4 \frac{\lambda}{(4\pi)^2} \left( 3 - 9 \xi - 3 \xi^2 - \alpha \right) + \bigo{\kappa^2} + \bigo{\lambda^2} \eqend{.}
\end{equations}

\subsection{Field redefinitions}

Recently~\cite{2martin2017} it has been suggested that the gravitational contribution to the $\beta$ function of the $\lambda \phi^4$ coupling could be cancelled by a field redefinition, similar to the quadratic terms in the $\beta$ function for Yang--Mills theory~\cite{ebertplefkarodigast2008,ellismavromatos2010}. Let us first review the argument for Yang--Mills theory, where the local part of the renormalised one-loop effective action has the form
\begin{equation}
\Gamma^\text{YM}_\text{loc} = - \frac{1}{2} \tr \int F^{\mu\nu} F_{\mu\nu} \total^4 x + d_1 \tr \int D_\rho F_{\mu\nu} D^\rho F^{\mu\nu} \total^4 x + d_2 \tr \int D_\mu F^{\mu\nu} D^\rho F_{\rho\nu} \total^4 x
\end{equation}
with higher-derivative couplings $d_1$ and $d_2$ (a possible third higher-derivative term proportional to $\tr F_{\mu\nu} F^{\nu\rho} F_\rho{}^\mu$ is equivalent to the first two up to a surface term). It turns out that at one-loop order, only the second higher-derivative term needs to be renormalized, and consequently only $d_2$ is renormalisation-scale dependent and we may set $d_1 = 0$. A (nonlinear) field redefinition
\begin{equation}
A_\mu \to A_\mu + \alpha D^\rho F_{\rho\mu}
\end{equation}
results, up to surface terms, in
\begin{equation}
\Gamma^\text{YM}_\text{loc} \to \Gamma_\text{YM} + 2 \alpha \tr \int D_\rho F^{\rho\nu} D^\mu F_{\mu\nu} \total^4 x + \bigo{\alpha^2} \eqend{,}
\end{equation}
and since $d_2$ is of order $\kappa^2$, we can remove the higher-derivative term to one-loop order by the choice $\alpha = - d_2/2$. There are no logarithmic contributions to the $\beta$ function (of the coupling $g$), which is clear from dimensional arguments since perturbatively Yang--Mills fields are massless and there is no mass scale with which $\kappa^2$ could be combined.\footnote{The same dimensional argument is of course valid for the scalar field, where in the massless case only the higher-derivative counterterm $\delta_{Z_1}$~\eqref{massless_higherder_ct} is non-vanishing, and the logarithmic contributions~\eqref{scalar_beta_functions} to the $\beta$ function vanish.} In total, it thus follows that all one-loop gravitational corrections can be absorbed into the above field redefinition, and there is no physical effect (e.g., the S-matrix is independent of the redefinition~\cite{weinberg_v1}). Note however that the vanishing of the logarithmic contributions at one-loop order is important; at two-loop order one expects a logarithmic contribution due to graviton loops~\cite{rodigast2012} which cannot be removed by a field redefinition.

We now try similarly to remove the higher-derivative terms in the effective action for scalar fields. The local part of the renormalised effective action reads
\begin{equation}
\Gamma_\text{loc} = - \frac{1}{2} \int \left[ \partial^\mu \phi \partial_\mu \phi + m^2 \phi^2 + \kappa^2 \delta_{Z_1}^\text{fin} \left( \partial^2 \phi \right)^2 \right] \total^4 x - \frac{\lambda}{4!} \int \phi^4 \total^4 x - \frac{\kappa^2}{4!} \delta_{Z_2}^\text{fin} \int \phi^3 \partial^2 \phi \total^4 x \eqend{,}
\end{equation}
and the nonlinear field redefinition
\begin{equation}
\phi \to \phi + \alpha_1 \kappa^2 \phi^3 + \alpha_2 \kappa^2 \partial^2 \phi + \alpha_3 \kappa^2 m^2 \phi
\end{equation}
together with the mass and coupling constant redefinition
\begin{equation}
m^2 \to m^2 + \alpha_4 \kappa^2 m^4 \eqend{,} \qquad \lambda \to \lambda + \alpha_5 \kappa^2 m^2
\end{equation}
leads after a straightforward calculation to
\begin{splitequation}
\Gamma_\text{loc} &\to - \frac{1}{2} \left( 1 - 2 \alpha_2 \kappa^2 m^2 + 2 \alpha_3 \kappa^2 m^2 \right) \int \left( \partial^\mu \phi \partial_\mu \phi \right) \total^4 x \\
&\quad- \frac{1}{2} \left( 1 + 2 \alpha_3 \kappa^2 m^2 + \alpha_4 \kappa^2 m^2 \right) \int m^2 \phi^2 \total^4 x - \frac{\kappa^2}{2} \left( \delta_{Z_1}^\text{fin} - 2 \alpha_2 \right) \int \left( \partial^2 \phi \right)^2 \total^4 x \\
&\quad- \frac{\lambda}{4!} \left( 1 + \frac{4!}{\lambda} \alpha_1 \kappa^2 m^2 + 4 \alpha_3 \kappa^2 m^2 + \frac{1}{\lambda} \alpha_5 \kappa^2 m^2 \right) \int \phi^4 \total^4 x \\
&\quad- \frac{\kappa^2}{4!} \left( \delta_{Z_2}^\text{fin} + 4 \alpha_2 \lambda - 4! \alpha_1 \right) \int \phi^3 \partial^2 \phi \total^4 x - \frac{\lambda}{3!} \alpha_1 \kappa^2 \int \phi^6 \total^4 x + \bigo{\kappa^4} \eqend{.}
\end{splitequation}
In order to maintain the field strength, mass and coupling constant normalisation we must choose
\begin{equation}
\alpha_3 = \alpha_2 \eqend{,} \qquad \alpha_4 = - 2 \alpha_2 \eqend{,} \qquad \alpha_5 = - 4 \alpha_2 \lambda - 4! \alpha_1 \eqend{.}
\end{equation}
The higher-derivative terms can be removed by the choice
\begin{equation}
\alpha_2 = \frac{1}{2} \delta_{Z_1}^\text{fin} \eqend{,} \qquad \alpha_1 = \frac{1}{4!} \delta_{Z_2}^\text{fin} + \frac{2 \lambda}{4!} \delta_{Z_1}^\text{fin} \eqend{,}
\end{equation}
and we obtain
\begin{equation}
\Gamma_\text{loc} = - \frac{1}{2} \int \left( \partial^\mu \phi \partial_\mu \phi + m^2 \phi^2 \right) \total^4 x - \frac{\lambda}{4!} \int \phi^4 \total^4 x - \frac{\lambda \kappa^2}{144} \delta_{Z_2}^\text{fin} \int \phi^6 \total^4 x + \bigo{\kappa^4} + \bigo{\lambda^2} \eqend{.}
\end{equation}
While the higher-derivative terms have vanished, a new (effective, non-renormalisable) operator $\phi^6$ has appeared. It is thus seen to be impossible to completely remove such operators, and in this sense the corrections are physical~\cite{anberdonoghueelhoussieny2011}. However, since in the massless case only $\delta_{Z_1}$ receives a divergent contribution we may set $\delta_{Z_2} = 0$ in this case. Only the field redefinition
\begin{equation}
\phi \to \phi + \frac{1}{2} \delta_{Z_1}^\text{fin} \kappa^2 \partial^2 \phi
\end{equation}
is then necessary, and the redefined effective action does not depend on higher-derivative terms in its local part. Nevertheless, unlike the Yang--Mills case there are logarithmic corrections to the $\beta$ function in the massive case. Under the above redefinition, we have
\begin{splitequation}
\beta &\to \beta - \kappa^2 m^2 \left( 4 \lambda \, \mu \frac{\partial}{\partial \mu} \delta_{Z_1}^\text{fin} + \mu \frac{\partial}{\partial \mu} \delta_{Z_2}^\text{fin} \right) - \kappa^2 \left( 4 \lambda \, \delta_{Z_1}^\text{fin} + \delta_{Z_2}^\text{fin} \right) \left( \mu^2 \beta_m + 2 m^2 \right) \\
&= - \frac{8 \lambda \kappa^2}{(4\pi)^2} m^2 \left[ 3 - 14 \xi + 9 \xi^2 - \alpha \right] + \bigo{\kappa^4} + \bigo{\lambda^2} \eqend{,}
\end{splitequation}
and we see that in general it is impossible to make the $\beta$ function vanish.

The authors of~\cite{2martin2017} only study the minimally coupled case $\xi = 0$, and make the (re)definition (converted to our normalisation, which changes $\kappa^2$ by a factor of $2$ and the gauge parameter $\alpha$ by a factor of $-2$)
\begin{equation}
\label{martin_redefinition}
\phi_0 = \phi - \frac{1}{4 \pi^2 (n-4)} \kappa^2 m^2 \phi + \frac{(3-\alpha)}{384 \pi^4 (n-4)^2} \kappa^2 m^2 \lambda \kappa^2 \mu^{-(n-4)} \phi^3 \eqend{,}
\end{equation}
where $\phi_0 = \sqrt{Z} \phi$ is the bare field. In the minimally coupled case, we have $Z = 1 + \delta_Z$ with~\eqref{coupling_counter2}
\begin{equation}
\delta_Z = \frac{2}{n-4} \mu^{n-4} \frac{\kappa^2}{(4\pi)^2} m^2 (3 - \alpha) \eqend{,}
\end{equation}
such that the result~\eqref{martin_redefinition} corresponds to the redefinition
\begin{equation}
\phi \to \phi - \frac{1}{2} \delta_Z \phi - \frac{1}{4 \pi^2 (n-4)} \kappa^2 m^2 \phi + \frac{(3-\alpha)}{384 \pi^4 (n-4)^2} \kappa^2 m^2 \lambda \kappa^2 \mu^{-(n-4)} \phi^3
\end{equation}
of the renormalised field. It is clearly seen that this is an infinite redefinition, which is not allowed in the renormalised theory (e.g., a similar infinite redefinition in quantum electrodynamics would remove the running of the electric charge, which has been experimentally measured). Moreover, the effective operator $\phi^6$ which appears in the action after such a redefinition would have an infinite coefficient, which is not treated at all by~\cite{2martin2017}. We conclude that it is impossible to remove the gravitational contributions to the $\beta$ function of the $\lambda \phi^4$ coupling, contrary to their suggestion.

\section{Outlook}
\label{sec_discussion}

We have shown how one can construct a particular class of relational observables around flat space, where the field-dependent coordinates on which the dynamical fields are evaluated are constructed purely from the metric perturbations. This is a generalisation of the proposal of~\cite{brunettietal2016}, and has the advantage that the dynamical content of the theory is not altered, contrary to other existing proposals such as the Brown--Kucha{\v r} dust~\cite{brownkuchar1995}. The condition that we impose on the field-dependent coordinates is that they are harmonic with respect to the perturbed d'Alembertian, and reduce on the background to the usual Cartesian coordinates,
\begin{equation}
\tilde{\nabla}^2 \tilde{X}^\mu = 0 \eqend{,} \qquad \tilde{X}^\mu \big\rvert_\text{BG} = x^\mu \eqend{.}
\end{equation}
Since curvature is measured by second derivatives, the background relation $\partial^2 x^\mu = 0$ (i.e., the Cartesian coordinates are harmonic with respect to the flat d'Alembertian) can be interpreted as saying that coordinate lines are straight. Therefore, in a sense the field-dependent coordinates $\tilde{X}^\mu$ determine a coordinate system which is ``as straight as possible'' in the perturbed geometry. While usually one would think of constructing such a coordinate system using (perturbed) geodesics, using geodesics in the construction of invariant correlation functions leads to highly singular results~\cite{khavkine2011,bongakhavkine2013,froeb2017}, and does not seem very viable. Note that in contrast to other approaches to construct invariant observables at higher order, in particular in the context of cosmology~\cite{rigopoulos2004,acquavivaetal2003,rigopoulos2006,finellietal2006,nakamura2006,malik2009,prokopecweenink2013}, our construction is systematic to all orders and independent of the actual gravitational theory. That is, there are no ad-hoc constructions involved such that in principle it can be implemented in a computer algebra system, and since it does not make use of the gravitational field equations or any specific action, it is equally valid for Einstein--Hilbert gravity (with our without cosmological constant), conformal (Weyl-squared) gravity, or higher-derivative theories of gravity.

We have then calculated one-loop graviton corrections to an invariant scalar two-point function. It could be seen very clearly that and how the gauge-dependent terms cancel between the usual field-theoretic contributions and the coordinate corrections, and the result has the expected functional form at one loop. We have also determined the gravitational contributions to the running of couplings, which results in a negative $\beta$ function for the $\lambda \phi^4$ coupling for most values of the non-minimal coupling parameter $\xi$ of the scalar field to curvature ($\xi R \phi^2$), including minimal coupling $\xi = 0$ and conformal coupling $\xi = 1/6$. By a non-linear field redefinition, one can get rid of the higher-derivative terms in the effective action, but a new effective $\phi^6$ term appears. Furthermore, it is not possible to cancel the gravitational contribution to the $\beta$ function in this way, unlike in the Yang--Mills case --- except if the scalar is massless, where to one-loop order there are no gravitational contributions to the $\beta$ function to begin with. Note in particular that unlike~\cite{ebertplefkarodigast2008,toms2010,toms2011,felipeetal2011,tangwu2012,felipeetal2013}, but in accordance with the more recent~\cite{ellismavromatos2010,pietrykowski2013}, we do not interpret contributions from higher-derivative (counter-)terms as quadratic contributions to the $\beta$ function of the coupling, but --- exactly as they arise in the effective action --- as contributions to the running of the coupling of a higher-derivative (or higher-dimension) effective operator. Whether or how this effective operator contributes to a concrete physical process/scattering experiment is a completely separate question; see~\cite{anberdonoghueelhoussieny2011,anberdonoghue2012} for a discussion and an explanation why dimensional regularisation is sufficient.

Of course, the obtained results functionally depend on the field-dependent coordinates; i.e., our $\tilde{X}^\mu$ only represent one way of defining a sensible coordinate system in the perturbed geometry, and for certain experiments or to quantify corrections to certain observables it might be more useful to use a different construction. For example, quantities at fixed Brown--Kucha{\v r} coordinates correspond to results seen by a family of observers who are comoving with the dust. However, as said above the Brown--Kucha{\v r} dust changes the physical content of the theory, and the construction presented in this work is applicable to situations in which there is no privileged or distinguished observer or family of observers.

The present work only investigates the corresponding relational observables at one-loop order. While the generalisation to higher-spin fields is straightforward, higher loop corrections may present difficulties. The principal obstacle is the non-locality of the invariant field $\phi(\xx)$, which arises through the non-locality of the coordinate corrections~\eqref{construction_pert_xresult}. At one-loop order, we have seen that the gauge-dependent terms cancel if the Green's function~\eqref{construction_pert_green} which appears in the construction is the massless scalar Feynman propagator, and that the choice of generalised Landau gauge~\eqref{landau_nonlinear} makes all coordinate corrections vanish. Therefore, even though the invariant field $\phi(\xx)$ is in general nonlocal, in this gauge it becomes a local field. Since the regularised result is explicitly independent of the gauge, the correlation functions of $\phi(\xx)$ can be renormalised using the usual purely local counterterms in the action. It remains to see if these two conditions persist to higher loop order in order to show that our construction is viable.

From the results~\eqref{inv2pf_result}, \eqref{inv2pf_result_massless_coordinate} and~\eqref{scalar_beta_functions} one sees that any quantum gravitational corrections are suppressed by $\ell_\text{Pl}^2/r^2$, where $\ell_\text{Pl}$ is the Planck length and $r$ is a characteristic length scale (for example, the Compton wavelength of the scalar particle). It is thus experimentally extremely difficult to actually measure these corrections, and in order to obtain some observable effects one should study corrections in other backgrounds, such as cosmological spacetimes. In fact, calculations of matter loop corrections to the gravitational potentials of a point particle in de~Sitter space~\cite{wangwoodard2015,parkprokopecwoodard2016,froebverdaguer2016,froebverdaguer2017} show that one can obtain corrections of the form $\ell_\text{Pl}^2 H^2 \ln(\mu r)$, where $H$ is the Hubble constant, $r$ the physical distance from the particle and $\mu$ the renormalisation scale. While the corrections are still extremely small at present times, during inflation they are small but non-negligible, and it has been conjectured that corrections from graviton loops could even grow in time. However, so far it was not possible to quantify the graviton loop corrections in an invariant way, and the present work could be a step towards obtaining an invariant result (using then, e.g., the original construction of~\cite{brunettietal2016}, or the one of~\cite{froeblima2017}).

\ack
It is a pleasure to thank Chris Fewster, Thomas-Paul Hack, Atsushi Higuchi, Nicola Pinamonti and Kasia Rejzner for discussions. This work is part of a project that has received funding from the European Union's Horizon 2020 research and innovation programme under the Marie Sk{\l}odowska-Curie grant agreement No. 702750 ``QLO-QG''.

I thank Jack Laiho and Kenny Ratliff for pointing out a missing factor of 2 in $\mathcal{I}^{(4,1)}$~\eqref{inv2pf_cali4}. A revision of the computation led to missing factors of 2 also in $\mathcal{I}^{(5,1)}$ and $\mathcal{I}^{(6,1)}$ and a missing term $4 (1-\alpha) m^2 [ 2 q^2 + k^2 - 2 (kq) ]$ in the numerator of $\tilde{J}^{(2)}$~\eqref{eq:tildej2}. These led to changes throughout the work including the results, but not the comparison with previous results (who took $\xi = 0$ or $m = 0$). The main conclusions remain unchanged. For a detailed corrigendum, see the corresponding \TeX\ file.

\appendix

\section{Integrals}
\label{app_integrals}

We determine some loop integrals. The massive tadpole is given by~\cite{smirnov2004}
\begin{splitequation}
\label{app_integrals_tadpole}
\int \frac{1}{q^2 + m^2 - \mathi 0} \frac{\total^n q}{(2\pi)^n} &= \mathi \frac{\Gamma\left( - \frac{n-2}{2} \right)}{(4\pi)^\frac{n}{2}} (m^2)^\frac{n-2}{2} \\
&= \frac{\mathi}{(4\pi)^2} m^2 \left[ \frac{2}{n-4} + \gamma - \ln (4\pi) + \ln m^2 - 1 \right] + \bigo{n-4} \eqend{,}
\end{splitequation}
while the massless tadpole (a scaleless integral) vanishes in dimensional regularisation. For a convolution with arbitrary powers of the propagators, we have the Feynman parameter representation~\cite{smirnov2004}
\begin{splitequation}
\label{app_integrals_convol}
&\int \frac{1}{[ q^2 + m_1^2 - \mathi 0 ]^{\lambda_1}} \frac{1}{[ (p-q)^2 + m_2^2 - \mathi 0 ]^{\lambda_2}} \frac{\total^n q}{(2\pi)^n} \\
&\quad= \mathi \frac{\Gamma\left( \lambda_1 + \lambda_2 - \frac{n}{2} \right)}{(4\pi)^\frac{n}{2} \Gamma(\lambda_1) \Gamma(\lambda_2)} \int_0^1 \frac{\xi^{\lambda_1-1} (1-\xi)^{\lambda_2-1}}{[ \xi (1-\xi) p^2 + \xi m_1^2 + (1-\xi) m_2^2 - \mathi 0 ]^{\lambda_1 + \lambda_2 - \frac{n}{2}}} \total \xi \eqend{.}
\end{splitequation}
We need this integral for $m_1^2 = 0$, $\lambda_1 = 1,2$ and $\lambda_2 = 1$, where we obtain
\begin{equation}
\label{app_integrals_convol1}
\int \frac{1}{q^2 - \mathi 0} \frac{1}{(p-q)^2 + m^2 - \mathi 0} \frac{\total^n q}{(2\pi)^n} = \mathi \frac{\Gamma\left( 2 - \frac{n}{2} \right)}{(4\pi)^\frac{n}{2}} \int_0^1 \frac{(1-\xi)^{\frac{n}{2}-2}}{[ \xi p^2 + m^2 - \mathi 0 ]^{2 - \frac{n}{2}}} \total \xi
\end{equation}
and
\begin{equation}
\label{app_integrals_convol2}
\int \frac{1}{[ q^2 - \mathi 0 ]^2} \frac{1}{(p-q)^2 + m^2 - \mathi 0} \frac{\total^n q}{(2\pi)^n} = \mathi \frac{\Gamma\left( 3 - \frac{n}{2} \right)}{(4\pi)^\frac{n}{2}} \int_0^1 \frac{\xi (1-\xi)^{\frac{n}{2}-3}}{[ \xi p^2 + m^2 - \mathi 0 ]^{3 - \frac{n}{2}}} \total \xi \eqend{.}
\end{equation}
The divergences in the first integral~\eqref{app_integrals_convol1} as $n \to 4$ come from the $\Gamma$ function in front, and expanding and performing the $\xi$ integral we obtain
\begin{splitequation}
\label{app_integrals_convol1b}
\int \frac{1}{q^2 - \mathi 0} \frac{1}{(p-q)^2 + m^2 - \mathi 0} \frac{\total^n q}{(2\pi)^n} &= - \frac{\mathi}{(4\pi)^2} \bigg[ \frac{2}{n-4} + \gamma - \ln (4\pi) + \ln m^2 - 2 \\
&\qquad+ \left( 1 + \frac{m^2}{p^2 - \mathi 0} \right) \ln\left( 1 + \frac{p^2 - \mathi 0}{m^2} \right) \bigg] + \bigo{n-4} \eqend{.}
\end{splitequation}
In the second integral~\eqref{app_integrals_convol2}, the divergences as $n \to 4$ come from a non-integrable singularity at $\xi = 1$. To extract it, we add an intelligent zero, writing
\begin{splitequation}
&\int \frac{1}{[ q^2 - \mathi 0 ]^2} \frac{1}{(p-q)^2 + m^2 - \mathi 0} \frac{\total^n q}{(2\pi)^n} = \mathi \frac{\Gamma\left( 3 - \frac{n}{2} \right)}{(4\pi)^\frac{n}{2}} \frac{1}{[ p^2 + m^2 - \mathi 0 ]^{3 - \frac{n}{2}}} \int_0^1 (1-\xi)^{\frac{n}{2}-3} \total \xi \\
&\qquad\qquad+ \mathi \frac{\Gamma\left( 3 - \frac{n}{2} \right)}{(4\pi)^\frac{n}{2}} \int_0^1 \left[ \frac{\xi (1-\xi)^{\frac{n}{2}-3}}{[ \xi p^2 + m^2 - \mathi 0 ]^{3 - \frac{n}{2}}} - \frac{(1-\xi)^{\frac{n}{2}-3}}{[ p^2 + m^2 - \mathi 0 ]^{3 - \frac{n}{2}}} \right] \total \xi \eqend{.}
\end{splitequation}
The first integral can be done exactly and then expanded around $n = 4$, while in the second we can set $n = 4$ and integrate. This gives
\begin{splitequation}
\label{app_integrals_convol2b}
&\int \frac{1}{[ q^2 - \mathi 0 ]^2} \frac{1}{(p-q)^2 + m^2 - \mathi 0} \frac{\total^n q}{(2\pi)^n} = \frac{1}{p^2 + m^2 - \mathi 0} \frac{\mathi}{(4\pi)^2} \\
&\qquad\quad\times \left[ \frac{2}{n-4} + \gamma - \ln (4\pi) + \ln m^2 + \left( 1 - \frac{m^2}{p^2 - \mathi 0} \right) \ln\left( 1 + \frac{p^2 - \mathi 0}{m^2} \right) \right] + \bigo{n-4} \eqend{.}
\end{splitequation}
We furthermore need the integral
\begin{splitequation}
\label{app_integrals_convol3}
\int \frac{q^2}{(p-q)^2 + m^2 - \mathi 0} \frac{\total^n q}{(2\pi)^n} &= \int \frac{p^2 - 2 (pq) + q^2}{q^2 + m^2 - \mathi 0} \frac{\total^n q}{(2\pi)^n} \\
&= \mathi (p^2-m^2) \frac{\Gamma\left( - \frac{n-2}{2} \right)}{(4\pi)^\frac{n}{2}} (m^2)^\frac{n-2}{2} \eqend{,}
\end{splitequation}
where we used the rotational invariance of the $q$ integral and the fact that scaleless integrals vanish in dimensional regularisation~\cite{leibbrandt1975}. Finally, we also need
\begin{splitequation}
\int \frac{1}{[ q^2 + m^2 - \mathi 0 ]^2} \frac{\total^n q}{(2\pi)^n} &= \mathi \frac{\Gamma\left( - \frac{n-4}{2} \right)}{(4\pi)^\frac{n}{2}} (m^2)^\frac{n-4}{2} \\
&= - \frac{\mathi}{(4\pi)^2} \left[ \frac{2}{n-4} + \gamma - \ln (4\pi) + \ln m^2 \right] + \bigo{n-4} \eqend{.}
\end{splitequation}

\providecommand\newblock{\ }
\bibliography{literature}

\end{document}